\documentclass[aps,prb,twocolumn,letterpaper,superscriptaddress]{revtex4}
\usepackage{amsmath, amsthm}
\usepackage{keyval}
\usepackage{graphicx,latexsym}
\usepackage{dcolumn}
\usepackage{bm}
\usepackage{color}
\usepackage{ulem} 
\newcommand{\ket}[1]{|#1\rangle}
\newcommand{\bra}[1]{\langle#1|}

\newcommand{\avg}[1]{\langle #1 \rangle}
\DeclareMathOperator{\Tr}{Tr}

\definecolor{magenta}{rgb}{1,0,1}
\usepackage{color}

\begin{document}
\title{Bond-dependent slave-particle cluster theory based on density matrix expansion}
\author{Zheting Jin}
\affiliation{Department of Applied Physics, Yale University, New Haven, Connecticut 06511, USA}
\author{Sohrab Ismail-Beigi}
\affiliation{Department of Applied Physics, Yale University, New Haven, Connecticut 06511, USA}
\affiliation{Department of Physics, Yale University, New Haven, Connecticut 06511, USA}
\affiliation{Department of Mechanical Engineering and Materials Science, Yale University, New Haven, Connecticut 06511, USA}
\date{\today}
\begin{abstract}

Efficient and accurate computational methods for dealing with interacting electron problems on a lattice are of broad interest to the condensed matter community.  For interacting Hubbard models, we introduce a cluster slave-particle approach that provides significant computational savings with high accuracy for total energies, site occupancies, and interaction energies.  Compared to exact benchmarks using density matrix renormalization group for $d$-$p$ Hubbard models, our approach delivers accurate results using two to three orders of magnitude lower computational cost.  Our method is based on a novel slave-particle decomposition with an improved description of particle hoppings, and a new density matrix expansion method where the interacting lattice slave-particle problem is then turned into a set of overlapping real-space clusters which are solved self-consistently with appropriate physical matching constraints at shared lattice sites between clusters.

\end{abstract}
\maketitle
\section{Introduction}

One of the outstanding challenges in condensed matter physics is to find computationally efficient and simultaneously accurate methods to describe interacting electron systems for large lattices (e.g., crystalline materials).  For the case where localized electronic interactions dominate, the Hubbard model provides a specific theoretical model that can describe key aspects of many important materials such as superconductors, magnets, or metal-insulator systems.  Hence, a large amount of research effort has been expended in creating and improving methods for Hubbard systems.

Exact methods for Hubbard systems are limited to low dimensions or finite sizes.  The ground state of the one-dimensional (1D) half-filled single-orbital Hubbard model can be solved analytically using the Bethe ansatz \cite{bethe1931theorie, lieb1968absence}.  In addition, extremely high accurate results can be obtained using numerical methods including the density matrix renormalization group (DMRG) \cite{white1992density, schollwock2005density} and quantum Monte Carlos (QMC) \cite{motta2020ground}. 
However, for higher-dimensional models, only small lattices (or small fragments of lattices) can be solved by these numerical methods.  Hence, an outstanding challenge is to find methods that work well in low and high dimensions, and (at present) this requires making approximations. 

One approximate approach for the Hubbard model is the slave-particle method, also known as the slave or auxiliary or subsidiary boson method (e.g., our previously published Boson Subsidiary-Solver (BoSS) software \cite{georgescu2021boson}.  It was first proposed \cite{barnes1976new,barnes1977new} for analytical calculations in the infinite interaction limit and as an alternative to the Gutzwiller variational approach \cite{gutzwiller1963effect,gutzwiller1965correlation}.  It was then generalized to finite interactions by a functional integral approach based on the slave-particle representation \cite{kotliar1986new, raczkowski2006interplay, fresard1997interplay}.  Since this approach requires one auxiliary slave-particle for each possible electronic configuration, whose number grows exponentially with the number of degrees of freedom on each site, the required computations can become expensive for complex materials.  Hence, more economical slave-particle methods have been developed.  These representations describe the slave-particles via electron occupation numbers.  Different methods have been developed based on the degrees of freedom treated by the slave particles, such as the slave-rotor method \cite{florens2004slave, rau2011emergence} which can serve as an impurity solver \cite{florens2002quantum}, the slave-spin method \cite{de2005orbital} which is orbital and spin selective, and a generalized approach that includes the above two (and other variants) methods \cite{georgescu2015generalized}.  An auxiliary symmetry-breaking field approach \cite{georgescu2017symmetry} was then introduced to overcome difficulties in achieving spontaneous symmetry breaking in these approaches. 

In slave-particle methods, the interacting electron problem is decomposed into a non-interacting spinon problem on a lattice (easily solved using Bloch's theorem and diagonalization) and an interacting slave-particle problem on a lattice.  The most common approach for solving the latter has been to use a single-site approximation.  This is very similar in spirit to the local single-site approximation in dynamical mean-field theory (DMFT) \cite{georges1996dynamical}, a state-of-the-art method for (approximate) solutions to Hubbard models.  Single-site slave-particle approaches predict Mott transitions in high-dimensional systems very well \cite{ lechermann2007rotationally, ferrero2009pseudogap, isidori2009rotationally, yu2011mott, ko2011magnetism, lau2013theory}.  However, as a result of stronger fluctuations in low-dimensional systems, the single-site treatments can cause significant errors.  For example, the exact solution of the half-filled 1D Hubbard model \cite{lieb1968absence} has no Mott transition for any finite interaction strength $U$, but a false Mott transition is predicted in the single-site slave-particle theory and DMFT \cite{georgescu2017symmetry, lee2019rotationally, bolech2003cellular, zgid2012truncated}.  

For higher accuracy, one must go beyond the single-site approximation and consider a local cluster of interacting lattice sites.  Several cluster extension methods \cite{zhao2007self, hassan2010slave} were proposed for slave-particle problems based on a cluster mean-field approximation.  If one sets certain quasiparticle renormalization factors to unity, the cluster slave-particle theory can be simplified \cite{ayral2017dynamical, lee2019rotationally} to the density-matrix embedding theory (DMET) \cite{knizia2012density, knizia2013density}. 
However, setting the renormalization factors to unity leads to the appearance of the bare inter-site hopping in the embedding spinon Hamiltonian, and this  cannot reproduce interaction-induced band narrowing, e.g., as predicted by $GW$ or DMFT calculations \cite{sakuma2013electronic, tomczak2014asymmetry, kim2015nature, craco2019lda+}.  
In addition, all existing cluster slave-particle, cluster DMFT, and cluster DMET methods describe the cluster as a multi-site impurity connected to an averaged external bath.  Inevitably, some chemical bonds with large associated hoppings are approximated as intercluster hoppings between different fragments of the systems, and the cutting and modification of these bonds to form the clusters can cause large errors \cite{staar2013dynamical, hahner2020continuous, nagai2019smooth}.  Consequently, the finite cluster size effects lead to a trade-off between cluster size and errors in these cluster methods \cite{lee2019rotationally, hettler1998nonlocal, hettler2000dynamical, staar2013dynamical, hahner2020continuous}.  The finite size errors are even more difficult to control in higher dimensional systems since the cluster surface grows with cluster radius $r$ as $r^{d-1}$.

In this work, we introduce a novel slave-particle method that addresses many prior shortcomings.  We present a bond-dependent slave-particle theory along with a cluster decomposition based on a density matrix expansion.  It has the following key features: 

(i) The degrees of freedom involved in each slave bond are orbital+spin+bond, and the free parameters in the slave operators are designed so that all unphysical, particle non-conserving inter-site hoppings are forbidden (this was impossible in prior slave-particle approaches).  

(ii) Instead of coupling the interacting problem (site or cluster) to a mean-field bath, a density matrix expansion approach is used to reduce the interacting slave-particle lattice problem to a set of separate cluster problems solved under appropriate constraints.  The clusters overlap with each other to span the whole lattice, so that they connect via the interacting density matrices on the shared sites instead of via a mean-field bath.  

(iii) The resulting numerical method is highly efficient and parallelizable: all the benchmark tests on $d$-$p$ Hubbard models described below take on the order of one CPU minute or serial computation to complete on a modern commodity laptop computer.  In addition, for a general $d$-dimensional lattice and for a fixed cluster size, the computational cost only grows quadratically with the number of clusters in the whole system.  The separate clusters can be solved in parallel, so that generalizations to large systems, higher dimensions, and multiple orbitals per lattice site will have reasonable computational costs.  

%
%
%

In the following, we describe the theory and then present numerical results based on its implementation.  As we will see, our theory reproduces remarkably accurate results with low computational cost compared to our benchmark results either from exact diagonalization or DMRG.  

\color{black}

\section{The slave-bond representation}
\label{sec:repres}
In this section, we introduce our slave-bond representation and explain how it generalizes standard site-based slave-particle methods in a manner that allows one to avoid all unphysical particle-number-violating hopping processes in the slave-particle problem.  We   compare our method to more familiar existing methods at each step to allow for a clear comparison in the reader's mind.

We consider Hubbard Hamiltonians of the form
\begin{equation}
    \hat{H} = -\sum_{\alpha\beta} t_{\alpha\beta} \hat{c}^\dag_{\alpha} \hat{c}_{\beta} + \sum_{\alpha} \epsilon_{\alpha} \hat{n}_{\alpha} + \hat{H}^{\text{int}}\,,
    \label{eq:HHubbard}
\end{equation}
where Greek-letter indices $\alpha, \beta$ combine the site indices $i,j$, orbital indices $m,m^\prime$, and spin indices $\sigma, \sigma^\prime$ together, ranging over all sites, orbitals, and spins in the system, i.e., $\alpha \equiv im\sigma$.  The $\hat{c}_\alpha$ is a fermion annihilation operator removing an electron from localized state $\alpha$, and $\hat{n}_\alpha=\hat c_\alpha^\dag\hat c_\alpha$ is the fermion number counting operator for state $\alpha$.  The $t_{\alpha\beta}$ and $\epsilon_\alpha$ denote  hopping  and onsite energies, respectively.  The electron-electron interaction term $\hat H^{\text{int}}$ is the sum of local operators at each site $\hat H^{\text{int}}_i$, which in the simplest case are given by the classic Hubbard ``$U$'' form
\begin{equation}
    \hat{H}^{\text{int}} = \sum_i\hat{H}^{\text{int}}_i = \sum_{i,m} U_{im} \hat{n}_{im\uparrow} \hat{n}_{im\downarrow}\,,
\end{equation}
where each site and spatial orbital can, in principle, have its unique interaction strength $U_{im}$.  Additional local interactions that depend on the electron counts $\hat n_\alpha$ are completely straightforward to include requiring no change of formalism  \cite{georgescu2015generalized,georgescu2017symmetry}.  

The standard approaches for slave-particles  \cite{florens2004slave, rau2011emergence, florens2002quantum, de2005orbital, georgescu2015generalized, georgescu2017symmetry, zhao2007self, hassan2010slave, lanata2015phase} replace the physical electron operators on each site by a combination of a non-interacting fermion (called a spinon) and an interacting auxiliary or slave particle in a completely local manner:
\[
\hat{c}_{\alpha} \rightarrow \hat{f}_{\alpha} \hat{O}_{\alpha}\,,
\]
where $\hat f_\alpha$ is the non-interacting fermion annihilation operator, and $\hat O_\alpha$ is the lowering ladder operator for the slave particles.  Accordingly, the original physical electron Hilbert space is mapped onto a larger Hilbert space $\mathcal{H} \rightarrow \mathcal{H}_f \otimes \mathcal{H}_s$ where $\mathcal{H}_f$ and $\mathcal{H}_s$ are the Hilbert spaces of the spinons and slave particles, respectively.  The spinon Hilbert space $\mathcal{H}_f$, being a fermionic one, contains the same degrees of freedom as the original electron problem, but the physical modes for the slave particles vary based on the type of slave model chosen.  For example, if all degrees of freedom $\alpha$ are explicitly described in slave-particle Hilbert space, it is known as the ``slave-spin'' or ``spin+orbital'' method \cite{de2005orbital, georgescu2015generalized}.  

The eventual goal of any slave-particle method is to have the spinons carry the fermionic spin of the original electron, while the slave-particles carry the charge of the electron, and by decoupling them one has two easier problems to solve (see the next section).  However, at this stage, one is still considering an exact reformulation, so the spinons and slave particles always move together during a hopping process in a correlated manner.   Mathematically, it means that in the enlarged Hilbert space $\mathcal{H}_f \otimes \mathcal{H}_s$, there is a subset of physical states where the number of spinons and slave particles are equal for each state $\alpha$ which form a faithful one-to-one representation of the original states of the physical electrons $\ket{n_{\alpha}}$:
\begin{equation}
    \ket{n_{\alpha}} \rightarrow \ket{n_{\alpha}^f=n_{\alpha}; N_{\alpha}=n_{\alpha}}\,,
    \label{equ:physst}
\end{equation}
where $n_\alpha, n^f_\alpha, N_\alpha$ are the occupation numbers of the physical electron, spinon, and slaves, respectively, and can take the values of 0 or 1.  We will call these states the ``physical'' or ``number-matching'' states in what follows.

With this short review concluded, we now define our slave-bond formalism.  Our formalism lives in the same Hilbert space as before $\mathcal{H}_f \otimes \mathcal{H}_s$, and the novelty is in how we choose to define the slave-particle operators.  Our approach is based on using the full microscopic set of local quantum numbers  $\alpha\equiv im\sigma$ for the slave-particle description.  We take a pair of localized states $\alpha,\beta$ to define the bond $\alpha\beta$ with associated hopping operators $\hat c_\alpha^\dag \hat c_\beta$ which moves an electron from $\beta$ to $\alpha$.  We define our slave-bond representation by using the spinon fermionic operators $\hat f_\alpha$ as before but defining  slaves-particle operators that have a bond index $\avg{\alpha\beta}$, so the operator replacement is now given by 
\begin{equation}
    \hat{c}_{\alpha}^\dag\hat{c}_{\beta} \rightarrow \hat{f}_{\alpha}^\dag\hat{f}_{\beta} \hat{O}_{\alpha\avg{\alpha\beta}}^\dag \hat{O}_{\beta\avg{\alpha\beta}}\,.
    \label{equ:cc2ffOO}
\end{equation}
To make the action of   $\hat{f}_{\alpha}^\dag\hat{f}_{\beta} \hat{O}_{\alpha\avg{\alpha\beta}}^\dag \hat{O}_{\beta\avg{\alpha\beta}}$ on the physical states $\ket{n_{\alpha}^f=n_{\alpha}; N_{\alpha}=n_{\alpha}}$ identical to that of $\hat{c}_{\alpha}^\dag\hat{c}_{\beta}$ on the original electron states $\ket{n_{\alpha}}$, Appendix~\ref{app:commute} shows that the lowering operator $\hat{O}_{\alpha\avg{\alpha\beta}}$ must take the form 
\begin{equation}
    \hat{O}_{\alpha\avg{\alpha\beta}} = 
    \begin{pmatrix}
        0 & 1\\
        c_{\alpha\avg{\alpha\beta}} & 0  
    \end{pmatrix}
    \,,
    \label{equ:Omat}
\end{equation}
with the basis ordered as $\{\ket{N_{\alpha}=0}$, $ \ket{N_{\alpha}=1}\}$.  The number $c_{\alpha\avg{\alpha\beta}}$ is often called the ``gauge'', and its value is arbitrary at present since we are dealing with the exact problem with no approximations within the physical subspace.  The constraints determining its value will be described in Sec.~\ref{sec:decomposition}.   Appendix~\ref{app:commute} also shows that the form of Eq.~(\ref{equ:Omat}) also guarantees that anti-commutation relations are obeyed for the collective bond spinon+slave operator:
\begin{equation}
\left\{\hat{f}_{\beta} \hat{O}_{\beta\avg{\alpha\beta}}, \hat{f}_{\alpha}^\dag \hat{O}_{\alpha\avg{\alpha\beta}}^\dag \right\}
= \delta_{\alpha\beta}\,.
\label{equ:fOcommute}
\end{equation}
It is important to clarify that the individual site-based slave-particle lowering operators $\hat{O}_{\alpha\avg{\alpha\beta}}$ resemble but are {\it not} bosonic field operators. They are defined in such a way to obey  Eq.~(\ref{equ:fOcommute}).  Hence, our theory is a slave-particle theory and not a slave-boson theory.

\begin{figure}[t]
\begin{center}
\includegraphics[scale=0.6]{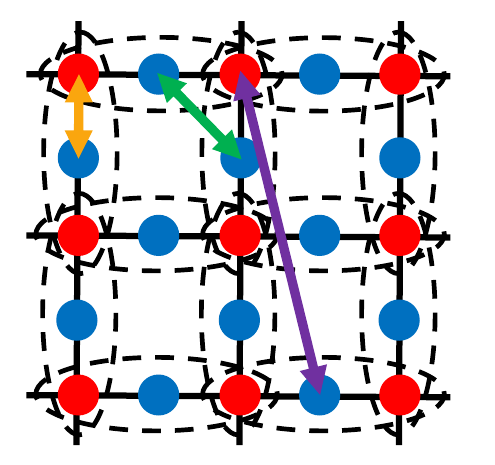}
\end{center}
\caption{
 An illustration of a checkerboard lattice structure representing a metal oxide 2D layer.  Red circles represent correlated $d$ sites (transition metals), and blue circles are non-interacting $p$ sites (oxygens).  The double-arrow lines illustrate examples of bonds used to define the slave-particle operators.  The dashed black ellipses indicate the $d$-$p$-$d$ clusters in the layer which overlap with each other on the correlated $d$ sites.
}
\label{fig:clusters}
\end{figure}

In contrast to prior slave-particle theories, the slave-bond operators in our theory are non-local.  Figure \ref{fig:clusters} shows examples of bonds (double-arrow lines) on a checkerboard lattice, where the bonds can correspond to hopping processes in the Hamiltonian, e.g., nearest-neighbor hopping (orange) and next-nearest-neighbor hopping (green).  However, one can also consider longer-ranged slave-bonds (purple).  Therefore, in principle, there are a huge number of bonds in a crystalline system.  But, in practice, only a small subset contributes to observables like the Hamiltonian.  For example, only bonds with non-zero hoppings $t_{\alpha\beta}\ne0$ contribute to the total energy, and these are typically only the nearest and next-nearest neighbors.  From a pragmatic viewpoint, in our work below we only need to define slave-bond operators on the bonds that appear with $t_{\alpha\beta}\ne0$ in Eq.~(\ref{eq:HHubbard}).

To compare to previous slave-particle methods,  for a given bond $\avg{\alpha\beta}$, our slave-bond approach can be viewed as a recipe similar to the prior site-based slave-particle approaches where one does the replacement
\begin{equation}
    \hat{c}_{\alpha} \rightarrow \hat{f}_{\alpha} \hat{O}_{\alpha\avg{\alpha\beta}}\,.
    \label{equ:c2fO}
\end{equation}
Even though the mapping of Eq.~\eqref{equ:c2fO} is mathematically correct (as detailed in Appendix~\ref{app:commute}), it shows an index mismatch from both sides.  Physically, it originates from the fact that the slave bonds are non-local; operationally, what it means is that mapping of Eq.~\eqref{equ:c2fO} only makes sense in the context of the hopping part of the Hamiltonian which is the sum over bonds: for each bond $\alpha\beta$, one can do the mapping in Eq.~\eqref{equ:c2fO} without confusion.

\section{Slave-particle Decomposition}
\label{sec:decomposition}
With the slave-bond representation, the Hamiltonian of Eq.~\eqref{eq:HHubbard} in the enlarged Hilbert space turns into 
\begin{equation}
\begin{split}
    \hat{H} = -\sum_{\alpha\beta} t_{\alpha\beta} \hat{f}^\dag_{\alpha} \hat{f}_{\beta} \hat{O}^\dag_{\alpha\avg{\alpha\beta}} \hat{O}_{\beta\avg{\alpha\beta}} + \sum_{\alpha} \epsilon_{\alpha} \hat{n}_{\alpha} 
    + \sum_i \hat{H}^{\text{int}}_i\,,
\end{split}
\end{equation}
where, again, $i$ is a site index and Greek-letter indices combine site, orbital, and spin together.  The local interaction term in the original electron Hamiltonian can be described by the slave particles alone through $\hat{H}^{\text{int}}_i = \sum_m U_{im} \hat{N}_{im\uparrow} \hat{N}_{im\downarrow}$.  The difficulty is that in addition to the physical or number-matching states, the enlarged Hilbert space $\mathcal{H}_f \otimes \mathcal{H}_s$ also contains numerous unphysical states, e.g.,  a state such as $\ket{n_{\alpha}^f=0; N_{\alpha}=1}$ where the number of spinons and slaves do not match for localized state $\alpha$.  In an exact treatment of the interacting problem, these states are excluded.  However, to make practical progress, one must make approximations.

The first approximation common to all slave-particle approaches, including ours, is to decouple the spinon and slave problems.  The simplest way forward is to approximate the density matrix for the joint spinon+slave system $\rho_{\text{tot}}$  by a product of a spinon density matrix $\rho_f$ and a slave density matrix $\rho_s$, i.e., $\rho_{\text{tot}}=\rho_f\otimes\rho_s$.  This decouples the two problems at the cost of losing the concerted or correlated description of the spinon and slave particles during the hopping process along each bond.  The best one can do is to ensure agreement on average.  Hence, one enforces a matching of the expectation values of the number operators for each state $\alpha$,
\begin{equation}
    \avg{\hat{n}_{\alpha}}_f = \avg{\hat{N}_{\alpha}}_s\,.
    \label{equ:constraint}
\end{equation}
Here, expectations are defined in the standard way: for any operator $\hat A$ acting in the spinon space, we have $\avg{\hat A}_f=\Tr(\hat A\hat\rho_f)$; similarly, for any operator $\hat B$ acting in the slave space, we have $\avg{\hat B}_f=\Tr(\hat B\hat\rho_s)$. Here $\hat{n}_{\alpha}$ and $\hat{N}_{\alpha}$ are number operators in spinon and slave Hilbert space, and we will drop the $f$ superscript on the spinon number operator going forward.

Given this approximate density matrix, the average of a hopping process along a bond in Eq.~(\ref{equ:cc2ffOO}) factorizes as
\begin{equation}
    \avg{\hat{f}_{\alpha}^\dag\hat{f}_{\beta} \hat{O}_{\alpha\avg{\alpha\beta}}^\dag \hat{O}_{\beta\avg{\alpha\beta}}} = \avg{\hat{f}_{\alpha}^\dag\hat{f}_{\beta}}_f \avg{\hat{O}_{\alpha\avg{\alpha\beta}}^\dag \hat{O}_{\beta\avg{\alpha\beta}}}_s\,.
    \label{equ:ffOOdecouple}
\end{equation}
Most generally, the expectation of the product operator $\hat A\hat B$ of a spinon operator $\hat A$ and slave operator $\hat B$ factorizes as
\begin{equation}
\begin{split}
\avg{\hat{A}\hat{B}} &= \Tr \left(\hat{A}\hat{B} \hat\rho_f\otimes\hat\rho_s\right) \\
&= \Tr \left(\hat{A} \hat\rho_f\right) \Tr \left(\hat{B} \hat\rho_s\right) = \avg{\hat{A}}_f\avg{\hat{B}}_s\,.
\end{split}
\end{equation}

The decoupling in Eq.~\eqref{equ:ffOOdecouple} results in two simpler problems to be solved instead of the original electron problem.   The easiest way to achieve this is to consider the variational problem of minimizing the total energy $E=\avg{H}$ under the occupation number constraints of Eq.~\eqref{equ:constraint} as well as more obvious constraints of the normalization of the density matrices $\Tr(\hat\rho_f)=\Tr(\hat\rho_s)=1$. Using the Lagrange multiplier approach, we consider the unconstrained minimization of the function $F$:
\begin{equation}
\begin{split}
    F = \,& \avg{\hat{H}} - \sum_{\alpha} h_{\alpha} \left(\nu_{\alpha}-\avg{\hat{N}_{\alpha}}_s\right)-\sum_{\alpha} h^\prime_{\alpha} \left(\nu_{\alpha}-\avg{\hat{n}_{\alpha}}_f\right) \\
    &- \epsilon_f\left(\Tr(\hat\rho_f)-1\right) - \epsilon_s\left(\Tr(\hat\rho_s)-1\right)\,.
\end{split}
\label{equ:F_wholesys}
\end{equation}
Here $h_{\alpha}, h^\prime_{\alpha}$ are Lagrange multipliers for the mean occupation numbers of the slaves and spinons, while  $\epsilon_f, \epsilon_s$ are the Lagrange multipliers for the normalization constraints.  For convenience, we have enforced occupation number matching via separate matching to target spinon occupancies $\nu_{\alpha}$, whose values are determined variationally \cite{georgescu2017symmetry}.

The differential of $F$ versus the two independent variables $\hat\rho_f$ and $\hat\rho_s$ takes the form
\[
dF = \Tr([\hat H_f-\epsilon_f\hat I]d\hat\rho_f) + \Tr([\hat H_s-\epsilon_s\hat I]d\hat\rho_s)\,,
\]
where the effective Hamiltonians for spinons ($\hat{H}_f$) and slaves ($\hat{H}_s$)  are 
\begin{equation}
\begin{split}
    \hat{H}_f = &-\sum_{\alpha\beta} t_{\alpha\beta} \avg{\hat{O}^\dag_{\alpha\avg{\alpha\beta}} \hat{O}_{\beta\avg{\alpha\beta}}}_s \hat{f}^\dag_{\alpha} \hat{f}_{\beta} 
    + \sum_{\alpha} \left(\epsilon_{\alpha} + h^\prime_{\alpha}\right) \hat{n}_{\alpha} \,.\\
    \hat{H}_s = &-\sum_{\alpha\beta} t_{\alpha\beta} \avg{\hat{f}^\dag_{\alpha} \hat{f}_{\beta}}_f \hat{O}^\dag_{\alpha\avg{\alpha\beta}} \hat{O}_{\beta\avg{\alpha\beta}} 
    + \sum_{\alpha} h_{\alpha} \hat{N}_{\alpha} + \sum_i \hat{H}^{\text{int}}_i\,.
\end{split}  
\label{equ:Hfs}
\end{equation}
Therefore, the original interacting electron problem is turned into a non-interacting spinon problem with symmetry-breaking field (onsite energies) $h^\prime_{\alpha}$ and an interacting slave problem with onsite energies $h_\alpha$.  Both Hamiltonian problems contain hopping terms that are renormalized by averages of the other problem, and the renormalization factors are to be determined self-consistently at the minimum energy configuration.  The expectation values of observables are then described by the minimizing spinon and slave density matrices.   

Since we have created an approximation to the problem, the choice of gauge numbers  $c_{\alpha\avg{\alpha\beta}}$ will now matter.
In previous slave-particle methods \cite{florens2004slave, de2005orbital, georgescu2015generalized}, the gauge is chosen to ensure that in the non-interacting limit, the spinon system alone will faithfully describe the original electron problem.  In these prior approaches, the purely local replacement $\hat{c}_{\alpha} \rightarrow \hat{f}_{\alpha} \hat{O}_{\alpha}$ goes hand in hand with a single-site slave-particle approximation where $\hat\rho_s$ is approximated as a product over single-site density matrices $\hat\rho_s^i$ as $\hat\rho_s\approx\bigotimes_i \hat\rho_s^i$.  Then the slave expectation over each bond factorizes as well $\avg{\hat{O}_{\alpha}^\dag \hat{O}_{\beta}}_s\approx \avg{\hat{O}_{\alpha}^\dag}_s \avg{\hat{O}_{\beta}}_s$. The gauge numbers $c_\alpha$ are then chosen to ensure $\avg{\hat{O}_{\alpha}}_s=1$ when $\hat H_{int}^i=0$, and thus in the non-interacting limit, the spinon Hamiltonian $\hat H_f$ of Eq.~\eqref{equ:Hfs} will become identical to the original electron Hamiltonian since the slave hopping renormalization factor $\avg{\hat{O}_{\alpha}^\dag \hat{O}_{\beta}}_s$ is replaced by unity.  As we explain below, this methodology permits unphysical particle-number-violating hopping process which creates significant errors in the total energy.  

One of the virtues of our bond-based approach is that it eliminates such unphysical processes.  Consider the original electron Hamiltonian: the hopping on a given bond $\avg{\alpha\beta}$ is described by the term $-t_{\alpha\beta}\hat{c}^\dag_{\alpha} \hat{c}_{\beta}+\text{H.c.}$.  This term is Hermitian and particle-conserving.  In the Fock space, its only non-zero matrix elements are between the two states $ \ket{n_{\alpha}=0; n_{\beta}=1}$ and $ \ket{n_{\alpha}=1; n_{\beta}=0}$.  Within the exact slave-particle description in the physical subspace, the same is true of the corresponding hopping operator $-t_{\alpha\beta}\hat{f}^\dag_{\alpha} \hat{f}_{\beta} \hat{O}^\dag_{\alpha\avg{\alpha\beta}} \hat{O}_{\beta\avg{\alpha\beta}} + \text{H.c.}$ because the fermionic spinon part $\hat{f}^\dag_{\alpha} \hat{f}_{\beta}$ alone can ensure that all other matrix elements are zero.
However, once we approximately separate the spinon and slave problems, the slave-particle Hamiltonian $\hat H_s$ of Eq.~\eqref{equ:Hfs} does not 
necessarily conserve particle number.  The hopping along bond $\avg{\alpha\beta}$ in $\hat H_s$ is proportional to
\begin{equation}
    \hat{O}_{\alpha\avg{\alpha\beta}}^\dag \hat{O}_{\beta\avg{\alpha\beta}}+\text{H.c.}  = 
    \begin{pmatrix}
        0 & 0 & 0 & v^* \\
        0 & 0 & u^* & 0 \\
        0 & u & 0 & 0 \\
        v & 0 & 0 & 0 
    \end{pmatrix}
    \,,
\label{equ:slavehopping}
\end{equation}
where the basis is ordered as $\{\ket{N_{\alpha}=0; N_{\beta}=0}$, $ \ket{N_{\alpha}=0; N_{\beta}=1}$, $ \ket{N_{\alpha}=1; N_{\beta}=0}$, $\ket{N_{\alpha}=1; N_{\beta}=1}\}$, 
and $v = c_{\alpha\avg{\alpha\beta}}+c_{\beta\avg{\alpha\beta}}$ and $u = 1 + c_{\alpha\avg{\alpha\beta}}c_{\beta\avg{\alpha\beta}}^*$.  The physical processes are proportional to $u$ while the unphysical ones are proportional to $v$.  In contrast to prior slave-particle methods where the gauge $c_\alpha$ is a fixed number for each local slave mode $\alpha$, our approach provides the additional bond index which allows us to require the additional constraint $v=c_{\alpha\avg{\alpha\beta}} + c_{\beta\avg{\alpha\beta}}=0$ or $c_{\alpha\avg{\alpha\beta}}=-c_{\beta\avg{\alpha\beta}}$. And the remaining gauge freedom for bond $\avg{\alpha\beta}$ (i.e., the value of $c_{\alpha\avg{\alpha\beta}}$) is chosen to ensure the correct non-interacting limit for $\hat H_f$, namely that $\avg{\hat{O}^\dag_{\alpha\avg{\alpha\beta}} \hat{O}_{\beta\avg{\alpha\beta}}}_s=1$ at zero interaction strengths.  We take the gauge numbers $c_{\alpha\avg{\alpha\beta}}$ to be real so that the number of constraints (two) matches the number of gauge variables on each bond.  Detailed analytical formulas are summarized in Appendix~\ref{app:cgauge}. 

For a non-interacting problem, the slave renormalization factors are unity, so that the spinon Hamiltonian alone is sufficient to compute the total energy $\avg{\hat H}$ and match the original electron problem.  In addition, slave-particle methods are analytically exact at the large interaction limit \cite{barnes1976new,barnes1977new}, so one can view them as interpolation methods for finite interaction strengths.  

From a practical viewpoint, the spinon problem on the lattice is trivial to solve by diagonalizing the associated one-particle Hamiltonian matrix $\hat{H}^0$ given by $\hat{H}_{\alpha\beta}^0 = -t_{\alpha\beta}+\delta_{\alpha\beta}(\epsilon_\alpha+h'_\alpha)$ using Bloch's theorem.
However, the slave Hamiltonian is still an interacting many-body problem on a lattice and is impossible to solve exactly for a large lattice.  Some approximations are required as per the next section.

\section{Cluster approximation}
\label{sec:clusterapprox}
We now describe a novel cluster approximation based on a density matrix expansion that maps the infinite lattice slave problem onto a set of coupled finite-sized interacting clusters.  We use overlapping clusters so that each chemical bond, e.g., transition-metal-oxygen bond, is included in some clusters and will be described explicitly.  A choice of overlapping clusters for a two-dimensional corner-sharing (checkerboard) metal oxide layer is illustrated by the dashed black ellipses in Fig.~\ref{fig:clusters} where each cluster consists of three sites: two interacting $d$ sites and the $p$ site between them.

\begin{figure}[t]
\begin{center}
\includegraphics[scale=0.5]{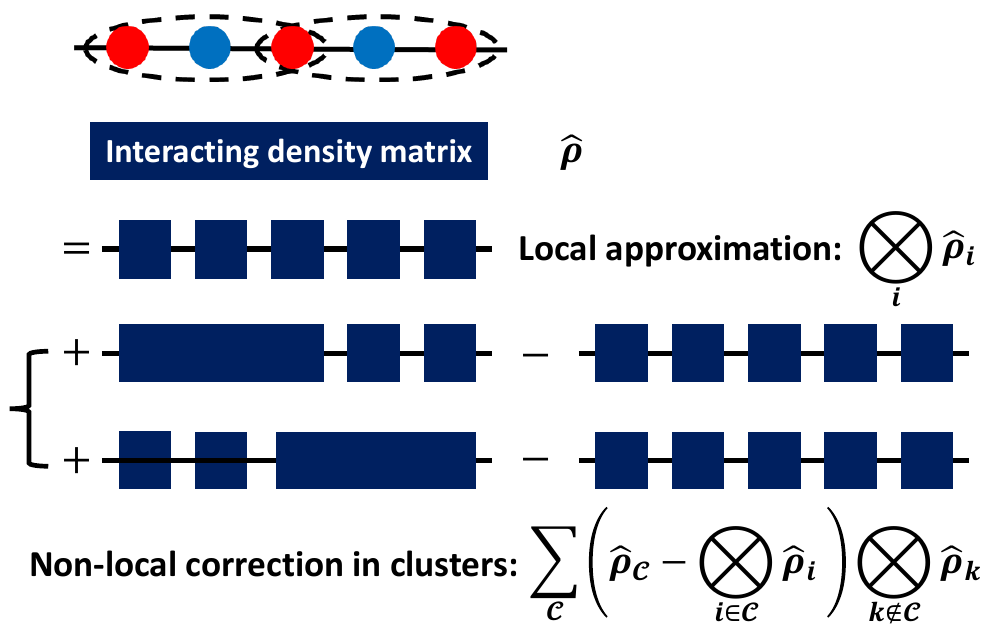}
\end{center}
\caption{
 An illustration of density matrix expansion for a one-dimensional system with red and blue atoms, where black dashed clusters overlap at red atom sites.  The interacting density matrix and its reduced density matrices on sites are shown as dark blue rectangles.  The direct product of density matrices is represented by black solid lines connecting rectangles. 
}
\label{fig:expand}
\end{figure}

Each cluster $\mathcal{C}$ is small enough that one can, in principle, solve the many-body interacting slave problem for that cluster directly.  For describing ground states, this would provide us with the cluster density matrix $\hat\rho_\mathcal{C}$. The question we address is how to take the set of $\{\hat\rho_\mathcal{C}\}$ over all the overlapping clusters and create a global quantum state (density matrix) for the entire lattice which we can then use to compute observables like the total energy.

The density matrix $\hat\rho$ of the entire  interacting slave problem will be approximated via the following real-space, site-based cluster expansion: 
\begin{equation}
    \hat\rho =  \bigotimes_i \hat\rho_i
    + \sum_\mathcal{C}\left(\hat\rho_\mathcal{C} - \bigotimes_{i \in \mathcal{C}} \hat\rho_i\right) \bigotimes_{k\notin \mathcal{C}}\hat\rho_k \,.
\label{equ:expan}
\end{equation}

The indices $i$ and $k$ refer to sites (atoms) in the system, and $\hat\rho_i$ is the single-site density matrix for site $i$ obtained by tracing out the degrees of freedom at all other sites
\begin{equation}
\hat\rho_i \equiv \Tr_{j\ne i}(\hat\rho)\,.
\label{equ:rhoidef}
\end{equation}
The first term in $\hat\rho$ of Eq.~\eqref{equ:expan} approximates the density matrix of the entire system as the tensor product over the single-site density matrices, and this represents the complexity of almost all current slave-particle approaches: each site $i$ is solved separately from the rest (albeit self-consistently via some type of bath linking the sites).  The second term in Eq.~\eqref{equ:expan} improves by incorporating the additional correlations described by the cluster compared to the single-site approximation.  
Figure \ref{fig:expand} illustrates the density matrix expansion of Eq.~\eqref{equ:expan}.  One begins with the collection of single-site density matrices $\hat\rho_i$ creating an approximation which is then improved by adding the cluster-wide density matrix contributions beyond the single-site approximation.  This expansion is trivially exact for an infinitely large cluster, while for finite-sized clusters, spatial correlations up to the cluster size are explicitly included.  
These clusters connect or handshake with each other via the single-site density matrices $\hat\rho_i$ on the shared sites as explained below.  

Since the $\hat\rho_{\mathcal C}$ are the basic variables describing the full system $\hat\rho$, the single-site $\hat\rho_i$ of Eq.~\eqref{equ:rhoidef} must also be derivable from the $\hat\rho_\mathcal{C}$.  Inserting Eq.~\eqref{equ:expan} into Eq.~\eqref{equ:rhoidef} yields the consistency condition
\begin{equation}
    \hat\rho_i =  \frac{1}{M_i}\sum_{\mathcal{C}|i\in\mathcal{C}} \Tr_{\mathcal{C}-i}(\hat\rho_\mathcal{C})\,,
    \label{equ:rhoiform}
\end{equation}
where $\Tr_{\mathcal{C}-i}(\hat\rho_\mathcal{C})$ is shorthand for 
\begin{equation}
\Tr_{\mathcal{C}-i}(\hat\rho_\mathcal{C}) \equiv \Tr_{k\in\mathcal{C}|k\neq i}(\hat\rho_\mathcal{C})
\end{equation}
which is the trace over all sites in cluster $\mathcal{C}$ excluding site $i$.  The number 
$M_i$ is the number of clusters that include site $i$: e.g., a $d$ site in Fig.~\ref{fig:clusters} has $M_i=4$ whereas a $p$ site belongs to a single cluster so it has $M_i=1$.  We will also use the shorthand
\begin{equation}
    \hat\rho_i^{(\mathcal{C})} =\Tr_{\mathcal{C}-i}(\hat\rho_\mathcal{C})
\end{equation}
for the single-site density matrix at site $i$ coming from the cluster density matrix of cluster $\mathcal{C}$.

Eq.~\eqref{equ:rhoiform} states the sensible condition that $\hat\rho_i$ is the average over all single-site density matrices coming from the cluster that overlap at site $i$.  However, to describe a consistent quantum state specified by $\hat\rho$ for all sites, we require a stronger consistency condition where the density matrix at each site is well-defined and unique.  Namely, we insist on the additional constraints that
\begin{equation}
    \hat\rho_i = \hat\rho_i^{(\mathcal{C})} \ \ \  \forall \mathcal{C}|i\in\mathcal{C}
    \label{eq:rhoiconsistency}
\end{equation}
separately for all clusters $\mathcal{C}$ containing $i$: i.e., all the $M_i$ contributions in Eq.~\eqref{equ:rhoiform} are the same.  Mathematically, for site $i$ in cluster $\mathcal{C}$ we employ a matrix of Lagrange multipliers $\hat{\Lambda}_i^{(\mathcal{C})}$ with associated Lagrange multiplier term $\Tr(\hat{\Lambda}_i^{(\mathcal{C})} [\hat\rho_i-\hat\rho_i^{(\mathcal{C})}])$ to enforce the constraint.

Two additional properties of this constrained density matrix expansion are: (i) it has a consistent description of short-ranged density matrices from the total density matrix $\hat\rho$, so $\Tr_{k\notin\mathcal{C}}(\hat\rho) = \hat\rho_\mathcal{C}$ 
and $\Tr_{k\neq i}(\hat\rho) = \hat\rho_i$; and (ii) it approximates the long-range behavior of the true density by single-site products: when sites $i$ and $j$ are far enough apart that they are not both in a single cluster, then $\Tr_{k\neq i,j}(\hat\rho) = \hat\rho_i\otimes\hat\rho_j$.  

Using the density matrix of Eq.~\eqref{equ:expan}, the total energy $E=\Tr(\hat H [\hat\rho_f\otimes\hat\rho_s])$ turns into
\begin{widetext}
\begin{equation}
\begin{split}
    E =&\;\sum_{\alpha} \epsilon_{\alpha} \avg{\hat{n}_{\alpha}}_{\rho_f} + \sum_i  \avg{\hat{H}^{\text{int}}_i}_{\rho_{i}}
    -\sum_{\alpha\beta | \avg{\alpha\beta}\in\exists\mathcal{C}}
    t_{\alpha\beta} \avg{\hat{f}^\dag_{\alpha} \hat{f}_{\beta}}_{\rho_f} 
    \avg{\hat{O}^\dag_{\alpha\avg{\alpha\beta}} \hat{O}_{\beta\avg{\alpha\beta}}}_{\rho_\mathcal{C}}\\
    &\;-\sum_{\alpha\beta | \avg{\alpha\beta}\notin\forall\mathcal{C}} 
    t_{\alpha\beta} \avg{\hat{f}^\dag_{\alpha} \hat{f}_{\beta}}_{\rho_f}
    \avg{\hat{O}^\dag_{\alpha\avg{\alpha\beta}}}_{\rho_{i|\alpha\in i}} \avg{\hat{O}_{\beta\avg{\alpha\beta}}}_{\rho_{j|\beta\in j}}
    \,.
\end{split}
\label{equ:Etotcluster}
\end{equation}
In this formula, the hopping energy has two parts.  The first is an intracluster hopping term describing hopping between two states $\alpha,\beta$ that are both inside of a cluster $\mathcal{C}$ (notation $\avg{\alpha\beta}\in\exists\mathcal{C}$): we can compute the associated slave-hopping expectation directly using the cluster density matrix without any factorization approximation.  The second intercluster hopping term is for long-ranged hoppings between $\alpha,\beta$ when both are not in a single cluster: here the slave-hopping expectation factorizes into the product of two single-site averages, and the consistency condition \eqref{eq:rhoiconsistency} ensures that the single-site averages of the hopping operators are well defined. 

To minimize $E$ with the required constraints, we use the Lagrange multiplier approach and consider the minimization of the function $F$:
\begin{equation}
\begin{aligned}
    F =\;& E - \sum _{i} \sum_{\mathcal{C}|  {i}\in\mathcal{C}}\Tr _{i} \left( \hat{\Lambda}_i^{(\mathcal{C})} 
    \left\{\left[\frac{1}{M _{i}}
    \sum_{\mathcal{C^\prime}|  {i}\in\mathcal{C^\prime}}\hat\rho_i^{(\mathcal{C^\prime})}
    \right]  
    -\hat\rho_i^{(\mathcal{C})}
    \right\}\right) \\
     &- \sum _{\alpha} \sum_{\mathcal{C}|  {\alpha}\in\mathcal{C}} h_{\alpha}^{(\mathcal{C})} \left(\nu_{\alpha}-\avg{\hat{N}_{\alpha}}_{\rho_\mathcal{C}}\right) -\sum_{\alpha}h^\prime_{\alpha} \left(\nu_{\alpha}-\avg{\hat{n}_{\alpha}} _{\rho_f}\right) 
    - \epsilon_f\left(\avg{\hat{I}} _{\rho_f}-1\right) - \sum_\mathcal{C} \epsilon _{\mathcal{C}}\left(\avg{\hat{I}} _{\rho_\mathcal{C}}-1\right)\,.
\end{aligned}
\label{equ:clusterF}
\end{equation}
Each cluster has its own Lagrange multipliers $h_{\alpha}^{(\mathcal{C})}$ to enforce mean occupancy matching with the spinons $\avg{\hat{n}_{\alpha}}_f = \avg{\hat{N}_{\alpha}}_{\rho_\mathcal{C}}$, and $\epsilon_{\mathcal C}$ enforce that the trace of the cluster density matrices are unity. Interestingly, as proved in Appendix~\ref{app:MoreConstraints}, the additional constraints introduced by $\hat{\Lambda}_i^{(\mathcal{C})}$ turn out to be redundant for the Hamiltonians of interest here given that the mean particle numbers are already matched to the spinon occupancies $\avg{\hat{n}_{\alpha}}_f$ for a slave mode $\alpha$ at site $i$.  So we can set all the  $\hat{\Lambda}_i^{(\mathcal{C})}=0$ which is a significant simplification.   

Redoing the logic of the minimization problem for this function $F$ yields the spinon Hamiltonian $\hat H_f$ and slave Hamiltonian $\hat H_\mathcal{C}$ governing each cluster $\mathcal{C}$:
\begin{equation}
\begin{split}
    \hat{H}_f =&\; -\sum_{\alpha\beta} t_{\alpha\beta} \avg{\hat{O}^\dag_{\alpha\avg{\alpha\beta}} \hat{O}_{\beta\avg{\alpha\beta}}}_\rho \hat{f}^\dag_{\alpha} \hat{f}_{\beta} + \sum_{\alpha} \left(\epsilon_{\alpha} + h^\prime_{\alpha}\right) \hat{n}_{\alpha} \,.\\
    \hat{H}_\mathcal{C} =
    &\;\sum_{i|i\in\mathcal{C}} \frac{1}{M_i} \hat{H}^{\text{int}}_i
    +\sum_{\alpha|\alpha\in\mathcal{C}} h_{\alpha}^{(\mathcal{C})} \hat{N}_{\alpha}
    -\sum_{\alpha\beta | \avg{\alpha\beta}\in\mathcal{C}} t_{\alpha\beta} \avg{\hat{f}^\dag_{\alpha} \hat{f}_{\beta}}_{\rho_f} \hat{O}^\dag_{\alpha\avg{\alpha\beta}} \hat{O}_{\beta\avg{\alpha\beta}}\\
    &\;-\sum_{\substack{\beta | \beta\in\mathcal{C} \\ \alpha | \avg{\alpha\beta}\notin\forall\mathcal{C}^\prime}} \frac{t_{\alpha\beta}}{M_\beta} \avg{\hat{f}^\dag_{\alpha} \hat{f}_{\beta}}_{\rho_f}  \left[\avg{\hat{O}^\dag_{\alpha\avg{\alpha\beta}}}_{\rho_{i|\alpha\in i}} \hat{O}_{\beta\avg{\alpha\beta}}+\text{H.c.}\right]\,.
\end{split}  
\label{equ:Hfc}
\end{equation}
\end{widetext}
Here $M_\alpha$ is a shorthand that equals the number of clusters to which the state $\alpha$ belongs and is equivalent to $M_i$   for any $\alpha \in i$.  The scaling factors of $1/M_i$ and $1/M_\alpha$ in $\hat H_\mathcal{C}$  originate from the relations \eqref{equ:rhoidef} and \eqref{equ:rhoiform} for the single-site density matrix $\hat\rho_i$.    The cluster Hamiltonian $\hat H_\mathcal{C}$ contains intracluster hoppings (first hopping term) as well as long-range intercluster hoppings ($\avg{\alpha\beta}$ is not in any single cluster) going outside the cluster $\mathcal{C}$ whose renormalization factors depend on the spinons and also the other slave clusters.

Using the cluster approximation, the lattice slave-bond Hamiltonian $\hat{H}_s$ of Eq.~\eqref{equ:Hfs} is mapped into a set of clusters, each with its own Hamiltonian $\hat H_\mathcal{C}$.  The density matrix of the interacting slave-bond lattice problem is then described by the density matrices of the clusters from Eq.~\eqref{equ:expan}.  The cluster Hamiltonians $\hat H_\mathcal{C}$ in Eq.~\eqref{equ:Hfc} are, in general, not particle conserving due to the intercluster hopping terms that involve single raising and lowering $\hat O$ operators.  In addition, the mean number of particles on any site or over any cluster is, in general, not an integer for a $pd$ type model due to $d$-$p$ hybridization.  Hence, the cluster density matrix $\hat\rho_\mathcal{C}$ will describe a mixed state, automatically involving nearly degenerate low-energy states.  In practice, we describe these mixed states using a Boltzmann distribution for each cluster using a very small temperature.  In this work, we use exact diagonalization to solve each interacting cluster problem and then use the eigenstates to compute the Boltzmann distribution: this is not prohibitively difficult for the small three-site clusters we consider here.  For larger clusters or more complex systems, more efficient methods of finding the Boltzmann distributed density matrix not involving naive diagonalization can be explored in the future.

\begin{figure}[t]
\begin{center}
\includegraphics[scale=0.45]{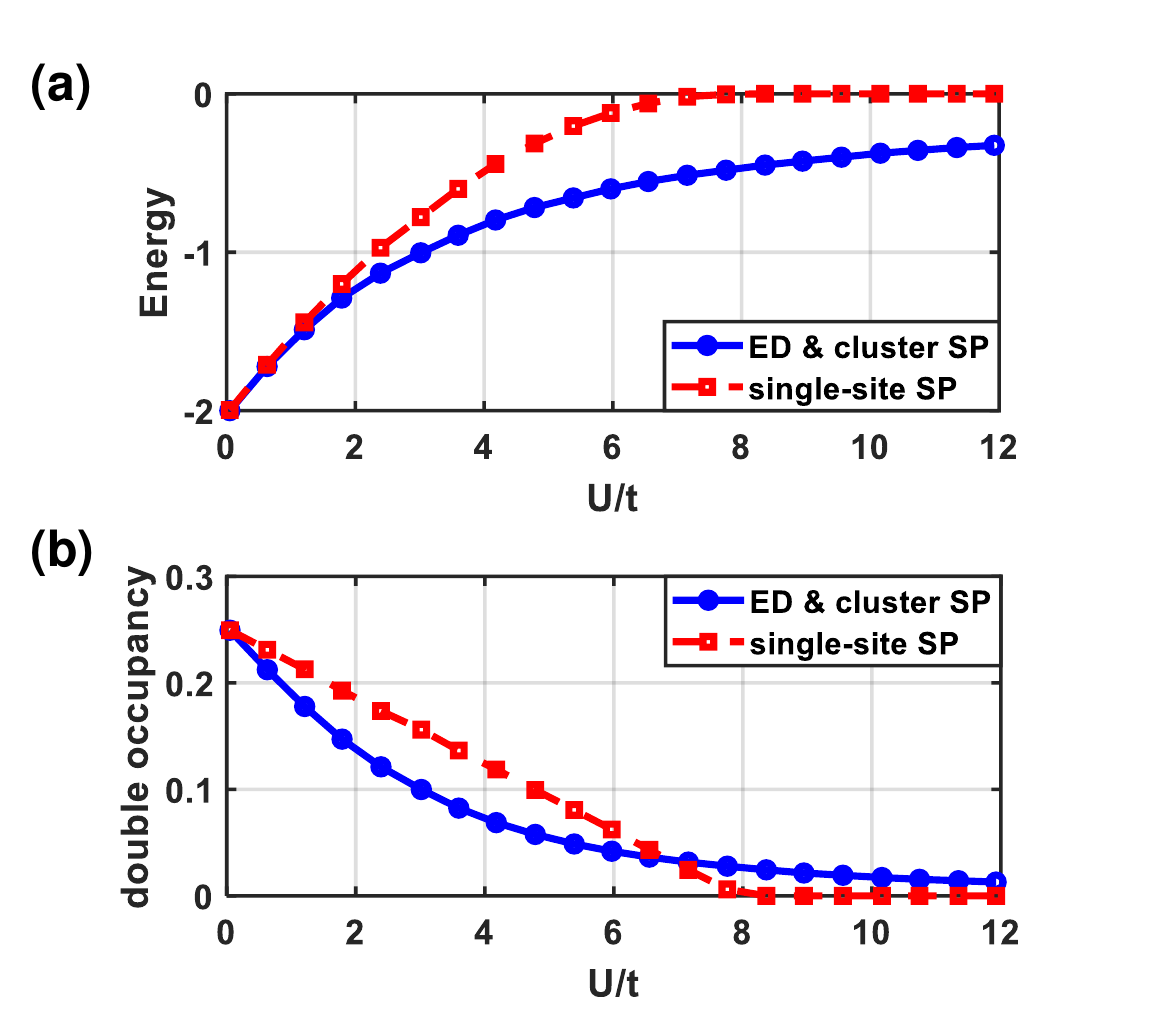}
\end{center}
\caption{
 (a) Total energy and (b) (mean) double occupancy of each site for the half-filled Hubbard dimer as a function of the interaction strength $U/t$, where the nearest-neighbor hopping strength $t$ is the energy unit.  Cluster slave-particle (SP) results are identical to the exact diagonalization (ED) results marked by blue circles, while single-site SP results are represented by red squares.  
}
\label{fig:Mott}
\end{figure}

\section{Tests on the Hubbard dimer}
\label{sec:Mott}
We begin with a simple system where the analytical solution of the electron problem as well as the single- and cluster-slave problems is possible.  We consider the half-filled two-site Hubbard model (Hubbard dimer), whose Hamiltonian is  
\begin{equation}
    \hat{H} = -t\sum_{\sigma}(\hat{c}_{1\sigma}^\dag \hat{c}_{2\sigma} + \hat{c}_{2\sigma}^\dag \hat{c}_{1\sigma}) + U \sum_{i=1}^2 \hat{N}_{i\uparrow}\hat{N}_{i\downarrow}
\end{equation}
For such a small system, a single two-site slave cluster encompasses the whole system, so it is not surprising that the cluster-based method will provide excellent results.  In fact, as detailed in Appendix~\ref{app:2site}, exact diagonalization of the original electron Hamiltonian and solution of our cluster slave-particle method result in the same ground-state energy $E = \left(U-\sqrt{U^2+16t^2}\right)/2$ as well as the same double occupancy 
\begin{equation}
    D_i~\equiv \avg{\hat{N}_{i\uparrow}\hat{N}_{i\downarrow}} 
    = \left(1-\frac{U}{\sqrt{U^2+16t^2}}\right)/4
\end{equation}
This double occupancy is always positive for any real-valued $U$ and $t$, direct evidence of the absence of a Mott transition in this system (also true of the one-dimensional half-filled one-band Hubbard model~\cite{lieb1968absence}).   Figure \ref{fig:Mott} shows the exact $E$ and $D_i$ with blue circles.

However, analytical solution (see Appendix~\ref{app:2site}) of the associated single-site slave-particle problem  results in $E = -(U-8t)^2/32t$ for $U/t<8$ and $E=0$ when $U/t\geq8$ and double occupancy $D_i = (8t-U)/32t$ when $U/t<8$ and $D_i=0$ when $U/t\geq8$.  Both energy and double occupancy indicate a Mott transition at $U/t=8$ within this (erroneous) single-site approximation.  Figure \ref{fig:Mott} shows the total energy $E$ and double occupancy $D_i$ of the single-site SP method with red squares.  This (false) Mott transition at $U/t=8$ disagrees with the exact solution but is also inevitable within a single-site approach: a single site connected to a bath described by a single expectation $\avg{O}_s$ cannot know if the bath is meant to describe a zero- or high-dimensional material problem; since a Mott transition can happen in higher-dimensional systems and is achievable in the single-site theoretical description when $\avg{O}_s=0$, it occurs for some sufficiently large $U/t$ in the single-site picture.

\section{Tests on $d$-$p$ models}
\label{sec:test1d}

Moving beyond the analytically solvable Hubbard dimer, we numerically test our theory on larger and more realistic models. In terms of modeling transition metal oxides, basic chemical considerations show that a minimal model should include both the $d$ orbitals of the transition-metal atoms and the $p$ orbitals of the oxygen atoms (a ``$d$-$p$'' model).  For example, the single-orbital-per-site $d$-$p$ model, also known as the Emery model \cite{zaanen1985band, emery1987theory, varma1987charge}, is intensively studied as a potential framework of the high-$T_c$ copper-based superconductors.  Due to the complexity caused by the explicit inclusion of the $p$ states, the $d$-$p$ model is numerically more challenging to solve but is also more realistic \cite{white2015doping, cui2020ground, chiciak2020magnetic} than  further simplified models such as the one-band Hubbard model \cite{anderson1987resonating} or the $t$-$J$ model \cite{zhang1988effective}.  
Hence, our numerical tests in this section will focus on the $d$-$p$ model, while   complementary one-band Hubbard model tests can be found in Appendix~\ref{app:testHubbard}.

For the systems studied below, the hopping renormalization factors and occupancies are determined self-consistently in a numerical fashion.  In Appendix~\ref{app:workflow}, we describe the workflow of the self-consistent calculations involved in our numerical studies.  As we will see below, our slave-bond method produces very accurate results compared to high-quality benchmark results in both 1D and 2D $d$-$p$ systems.

\begin{figure}[t]
\begin{center}
\includegraphics[scale=0.35]{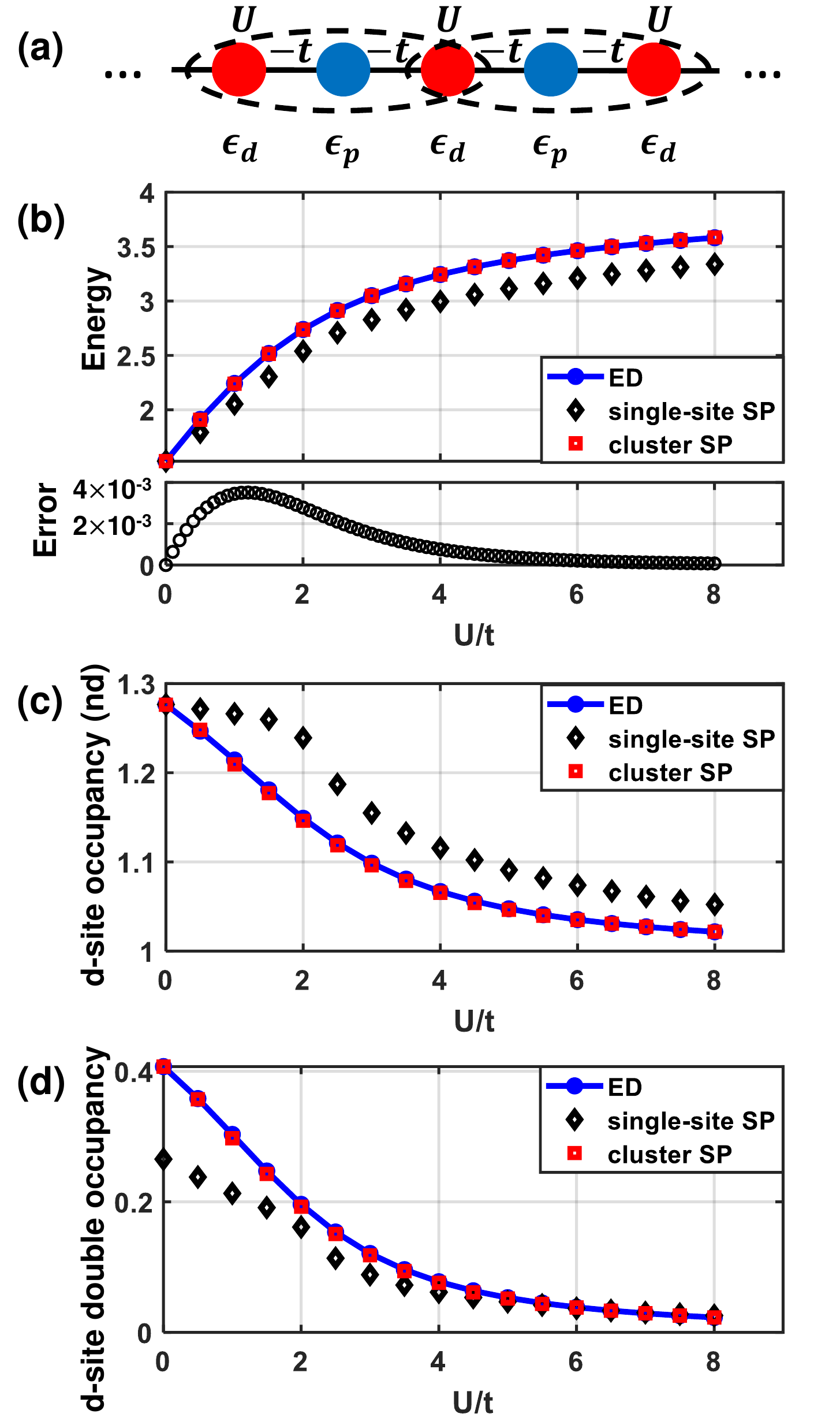}
\end{center}
\caption{
(a) Illustration of the one-dimensional $d$-$p$ chain with periodic boundary conditions, where red atoms represent $d$ sites while blue atoms represent $p$ sites.   The black dashed ellipses are the clusters used in the slave-bond calculation.   
Upper panel of (b), panels (c) and (d) show the total energy per unit cell in units of $t$, $d$-site occupancy, and $d$-site double occupancy versus the interaction strength $U$, respectively, for a four-site unit-cell linear chain with periodic boundary conditions.
The results of exact diagonalization (ED), single-site slave-particle (SP), and cluster SP are marked by blue circles, black diamonds, and red squares, respectively.  
The lower panel of (b) shows the total energy error of the cluster SP method in units of $t$. 
}
\label{fig:4site}
\end{figure}

We begin our tests with 1D $d$-$p$ systems which have the lattice illustrated in Fig.~\ref{fig:4site}(a).
In all the following results, the nearest-neighbor $d$-$p$ hopping strength is set to be $-t$ as illustrated in Fig.~\ref{fig:4site}(a), where $t$ is real positive and treated as the energy unit; all other hoppings are set to zero.  The onsite energy of $d$ sites is set to be $\epsilon_d = +2t$, while the onsite energy of $p$ sites is $\epsilon_p = 0$.  There is only one orbital for each site.  In addition, a small temperature of $k_BT = 5\times 10^{-3}t$ is introduced to allow the Boltzmann distribution to create a mixed state from multiple eigenfunctions of the Hamiltonian which is necessary as discussed  in Sec.~\ref{sec:clusterapprox}.  The finite temperature does create errors in the calculation of ground-state properties, but as shown in Appendix~\ref{app:temperature}, these errors are controllable by reducing the temperature.  Other convergence thresholds are set low enough to give accurate results: e.g., mean $d$-site occupancies are converged so that they differ between physically identical $d$ sites by less than $10^{-7}$.  

We employ two types of benchmarks.  For small systems, we use exact diagonalization (ED) of the starting electron Hamiltonian with a Boltzmann distribution at the same $k_BT$ listed above.  For larger problems, ED is infeasible, so we turn to DMRG.  All DMRG calculations in this work use the ITENSOR software package (Julia version) \cite{itensor, itensor-r0.3}, where the energy cutoff is set to be $10^{-8}t$; the maximum bond dimension increases by system size up to 1,600 for a 96-site system, and the maximum number of sweeps is 1,600.  For the single-site slave particle results, we use the Boson Subsidiary-Solver (BoSS) software \cite{georgescu2021boson} with orbital- and spin-resolved slave particles using the same finite temperature.  
\subsection{Four-site 1D model}
\label{sec:4site}
We begin with a small four-site model (i.e., a $d$-$p$-$d$-$p$ chain) with periodic boundary conditions (PBC). The upper panel of Fig.~\ref{fig:4site}(b) shows the total energy versus interaction strength $U$. Both  slave-particle (SP) methods reproduce exact total energy  at $U=0$ and $U\rightarrow\infty$ by construction.  
However, while the single-site method shows qualitatively correct behavior versus $U$, the cluster method shows quantitative accuracy.  The same is true for the mean $d$-site occupancy $n_d = \avg{\hat{N}_{d\uparrow}+\hat{N}_{d\downarrow}}$  in Fig.~\ref{fig:4site}(c).  The double-occupancy $D =\avg{\hat{N}_{d\uparrow}\hat{N}_{d\downarrow}}$ in Fig.~\ref{fig:4site}(d) is computed in the slave-particle sector of the problem which has no reason to match the exact answer at $U=0$, and the two SP calculations will match the exact answer only at $U\rightarrow\infty$.  However, the cluster SP method remains very accurate for all values of $U$. 

We note that for both energy and $d$-site occupancy, both cluster SP results agree with ED for both small and large $U$ which is as expected.  In the non-interacting limit, the $c$-gauges enforce the spinon Hamiltonian alone to reproduce the non-interacting electron Hamiltonian: with $\avg{\hat{O}^\dag_{\alpha\avg{\alpha\beta}} \hat{O}_{\beta\avg{\alpha\beta}}}=1$ and zero interaction terms, the total energy of Eq.~(\ref{equ:Etotcluster}) is the non-interacting energy.  In the large $U$ or atomic limit, the $p$ sites are filled while the $d$ sites are half-filled in both SP or ED calculations.  Thus, the cluster SP reproduces the exact energy and occupation numbers in both limits.  The single-site SP method has the same properties in both limits but has larger errors at finite $U$.  

The lower panel of Fig.~\ref{fig:4site}(b) shows the error of the total energy of the cluster SP as a function of interaction strength ($U/t$) calculated by subtracting the ED energy from the cluster SP energy.  To further reduce the finite temperature effect, the temperature is decreased to $k_BT = 1\times 10^{-4}t$ in this particular calculation.  Among all the 81 different interaction strengths $U/t$ sampled from 0 to 8, the maximum energy error in cluster SP is 3.5$\times10^{-3}t$ at $U/t=1.2$, while such error is about 50 times larger in the single-site SP method.  The maximum errors for occupancy and double occupancy in cluster SP are 1.1$\times10^{-3}$ and 1.6$\times10^{-3}$, respectively.

\subsection{System-size dependence in 1D}
\label{sec:systemsize}
\begin{figure}[t]
\begin{center}
\includegraphics[scale=0.4]{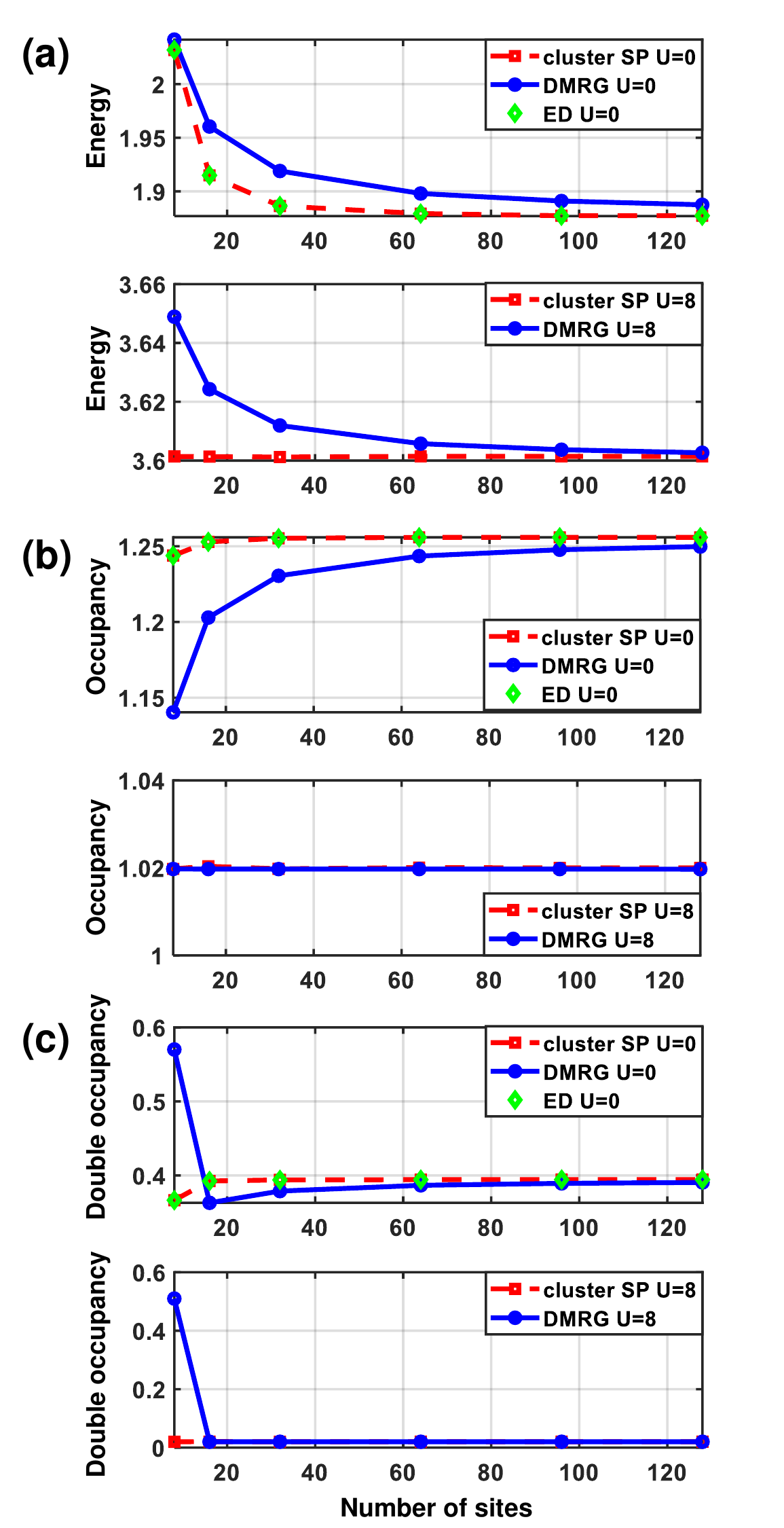}
\end{center}
\caption{
Convergence of observables versus system size for the 1D half-filled $d$-$p$ Hubbard chain: system sizes used are 8, 16, 32, 64, 96, and 128 sites.  (a)-(c) The total energy of each four-site unit cell in units of $t$, the $d$-site occupancy, and the $d$-site double occupancy versus the number of sites, respectively.  For each panel, the interaction strength used is $U=0$ for the upper sub-panel and $U=8t$ for the lower one.  The results of DMRG with OBC, cluster SP with PBC, and ED with PBC are marked by blue circles, red squares, and green diamonds, respectively.}  
\label{fig:diffnk}
\end{figure}
Beyond the four-site $d$-$p$ system where exact diagonalization is possible, we have tested longer 1D chains with 8, 16, 32, 64, 96, and 128 sites at $U/t =$ 0 and 8.  Here we assume translational symmetry with the four-site unit cell which is then replicated: the objective is to gauge convergence versus system size.  Benchmark results on the different-sized systems are generated by either exact diagonalization at $U=0$ or DMRG \cite{white1992density} with open boundary condition (OBC) at finite $U$.  We note that PBC is replaced by OBC in the DMRG calculations because PBC is computationally much more expensive than OBC according to the area law \cite{bauer2013area}.  Correspondingly, we calculate and report the  total energy per four-site unit cell in Fig.~\ref{fig:diffnk}(a) for OBC DMRG calculations to compare it with cluster SP and exact diagonalization results. The $d$-site occupancy and double occupancy in Figs.~\ref{fig:diffnk}(b) and \ref{fig:diffnk}(c) are computed around the middle point of the chain for DMRG so as to be the farthest away from the open boundaries.  
For the cluster SP calculations with PBC, we can use the translational symmetry to work with the four-site unit cell together with $k$-point sampling of the spinon problem: larger systems correspond to denser $k$ sampling of the non-interacting spinon problem. This makes for a very cheap computational scaling versus system size.

The system-size dependencies are shown in Fig.~\ref{fig:diffnk}, where the total energy is quoted per four-site unit cell.  
Unsurprisingly, the cluster SP method is equivalent to exact diagonalization at $U=0$ as discussed at the end of Sec. \ref{sec:decomposition}.  
We also notice that both cluster SP and DMRG converge faster versus system size at $U/t = 8$ case than at $U/t=0$.  This comes from the intensively studied ``band narrowing'' effect \cite{georgescu2015generalized, sakuma2013electronic, zhu2021ab} due to finite $U/t$ which makes the quasiparticle bands less dispersive and allows sparser $k$-sampling for the same accuracy.  In addition, cluster SP is well converged for most of the observables when the system size is larger than 16 sites, while DMRG needs more than one hundred sites for the same level of convergence.  The slower convergence of DMRG is due to edge effects induced by the open boundary condition needed to reduce bond dimension \cite{itensor, itensor-r0.3}. 

From a computational vantage point, 
to converge the total energy to $10^{-7}t$, each data point takes less than 1 CPU minute on a laptop for the slave SP calculations regardless of system size.  On the other hand, for the OBC DMRG, we require about 6 CPU hours for a 64-site calculation and takes longer than one day for a 128-site calculation on a standard, contemporary Linux cluster. 

\begin{figure}[t]
\begin{center}
\includegraphics[scale=0.35]{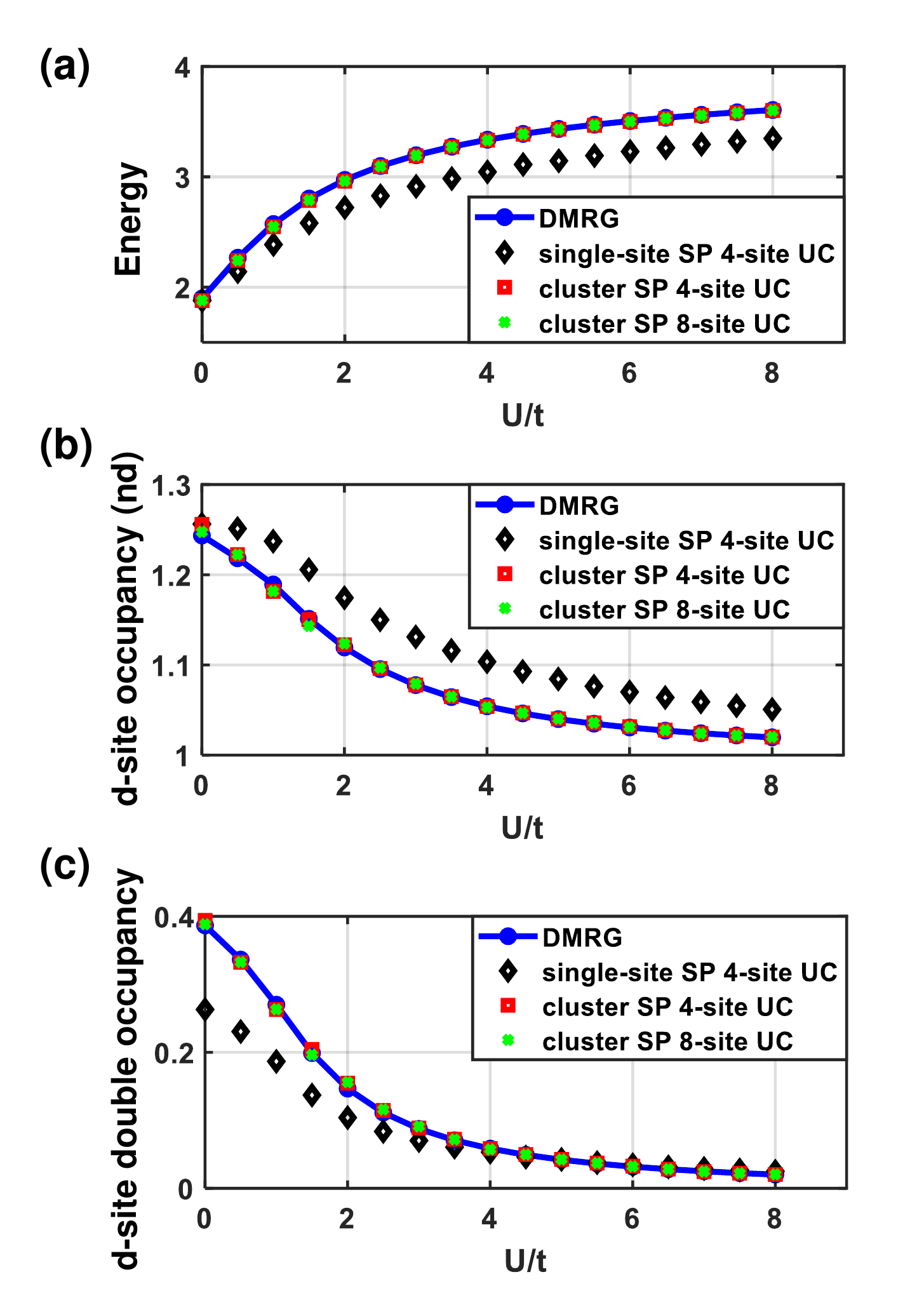}
\end{center}
\caption{
(a)-(c) The total energy of each four-site unit cell in units of $t$, $d$-site occupancy, and $d$-site double occupancy versus the interaction strength $U$, respectively.  
All results come from 64-site 1D $d$-$p$ system, blue circles represent DMRG results with OBC; black diamonds represent single-site SP results assuming a translational period of four sites; red squares and green crosses almost completely overlap with each other and represent cluster SP assuming a translational period of four sites and eight sites.  
}
\label{fig:64site}
\end{figure}

We further examine the 64-site system as a function of $U/t$ in Fig.~\ref{fig:64site}.   The cluster SP results marked by red squares assume a translational period of four sites and match well with the DMRG results marked by blue circles.  As a comparison, the single-site SP method assuming the same translational period is marked by black diamonds, which shows much larger errors in all three observables.  Moreover, an additional cluster SP calculation assuming a translation period of 8 sites is performed, whose results are marked by green crosses.  This calculation with a double-sized unit cell behaves almost exactly the same as the four-site unit cell calculation, which indicates the convergence of unit-cell size in our cluster SP results.  

\subsection{Doped 1D chains}
\label{sec:doping}
We perform further tests on the 64-site 1D $d$-$p$ system by hole doping, where the average doping density ranges from 0 up to 0.5 holes for each $d$-$p$ pair.  The local interaction strength on $d$ sites is fixed to be $U=2t$, while other parameters such as the onsite energies, hopping strengths, and the temperature are unchanged from the no-doping calculations of the previous sections.  We choose $U/t=2$ because the errors caused by the cluster SP method are relatively large around this choice, as shown in the lower panel of Fig.~\ref{fig:4site}(b), providing a stringent test of the cluster SP. 
\begin{figure}[t]
\begin{center}
\includegraphics[scale=0.35]{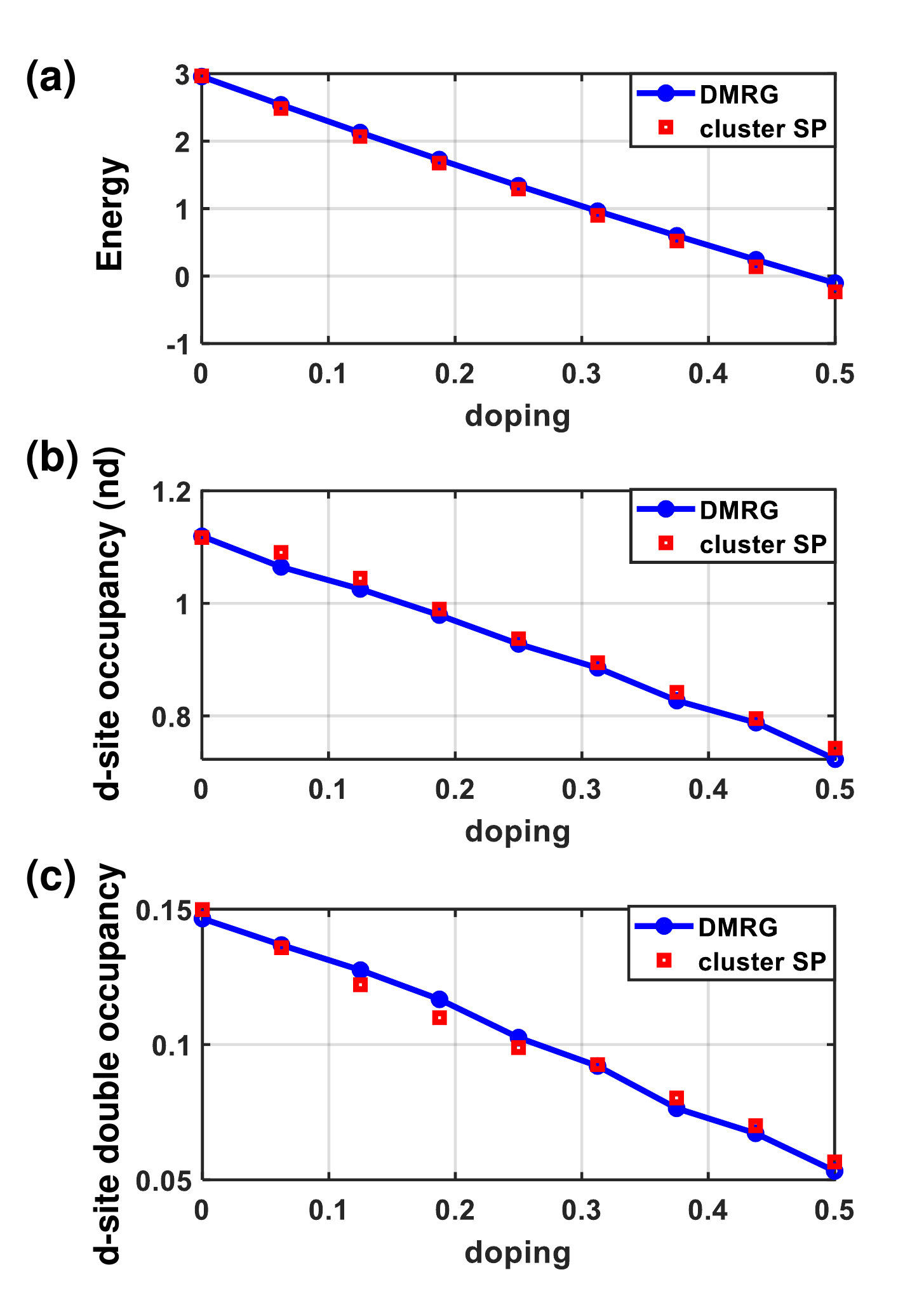}
\end{center}
\caption{
(a)-(c) The total energy of each four-site unit cell (units of $t$), $d$-site occupancy, and $d$-site double occupancy, respectively.  They are plotted as functions of the average hole doping level on each $d$-$p$ pair.  Blue circles represent the DMRG results as benchmarks, while the red squares stand for the cluster SP results assuming a translational period of four sites. 
}
\label{fig:doping}
\end{figure}

Our results of the doped systems are summarized in Fig.~\ref{fig:doping}, where the two curves are results from cluster SP and DMRG.  In Figs.~\ref{fig:doping}(a) and \ref{fig:doping}(b), the two curves are almost overlapping with each other, indicating that the cluster SP method reproduces the energy per unit cell and the $d$-site occupancy of the ground state extremely precisely.  In Fig.~\ref{fig:doping}(c), the cluster SP reproduces good $d$-site double occupancy $D$  with some small variations.  While both DMRG and cluster SP methods show a nice linear relation between the total energy and doping level in Fig.~\ref{fig:doping}(a), they both show small discrepancies away from linear relations in Figs.~\ref{fig:doping}(b) and \ref{fig:doping}(c).  This is because both DMRG and the cluster SP methods are variational approaches for the ground state total energy, so a high level of energy convergence is guaranteed, but the convergence of other non-variational observables such as the site occupancy or double occupancy is much poorer.  Note that the double-occupancy deviations are below 0.01, which is adequately small and about the same order of magnitude as other errors caused by the cluster approximation and finite-temperature effect and is almost invisible in Fig.~\ref{fig:64site}(c). 
 It is one order of magnitude smaller than the error caused by single-site SP approximation.   
 
In short, based on all these tests, our cluster SP method reproduces overall very accurate results hundreds of times faster than DMRG for the one-dimensional $d$-$p$ model.  It also represents a significant quantitative improvement over the single-site SP method.

\begin{figure}[t]
\begin{center}
\includegraphics[scale=0.33]{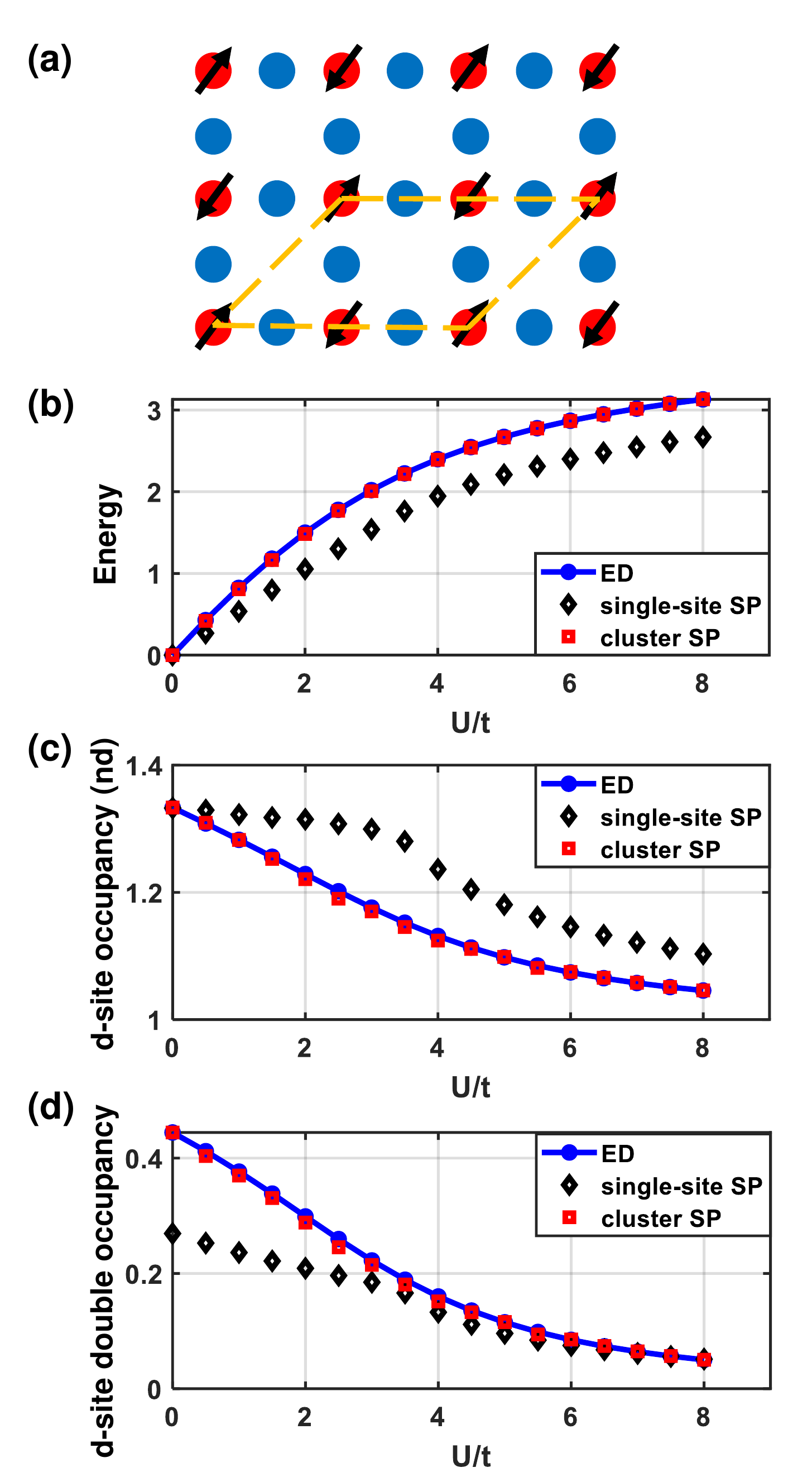}
\end{center}
\caption{
(a) Illustration of the purposed spin structure in a checkerboard lattice, where the nearly half-filled $d$ sites marked by red circles are spin-polarized, while the nearly full-filled $p$ sites marked by blue circles are not spin-polarized.  The orange dashed parallelogram represents the primitive cell under the N\'eel AFM correlation.  
(b)-(d) The total energy for each primitive cell, $d$-site occupancy, and $d$-site double occupancy versus the interaction strength $U$, respectively.  
The results of ED, single-site SP, and cluster SP methods are marked by blue circles, black diamonds, and red squares, respectively.  
}
\label{fig:checkerboard}
\end{figure}

\subsection{2D $d$-$p$ systems}
\label{sec:2D}
The generalization to two-dimensional systems is straightforward in our cluster slave-particle theory.  We choose a checkerboard lattice where the lattice structure and the clusters used in the calculations are illustrated in Fig.~\ref{fig:clusters}.  The red and blue circles represent $d$ and $p$ sites, where each site contains only one orbital.  This is a frequently studied model for cuprates known as the ``three-band model'' \cite{white2015doping, cui2020ground}.  

Figure \ref{fig:checkerboard}(a) illustrates a primitive cell used in the calculation which is a parallelogram and contains two $d$ sites and four $p$ sites.  This is because we are permitting for N\'eel checkerboard anti-ferromagnetic (AFM) correlation (if we assumed a stripe pattern, the unit cells would be chosen differently).  Similar to the 1D  calculations discussed above, the nearest-neighbor hopping strength $t$ is treated as the energy unit.  All the onsite energies of $d$ sites are set to be $\epsilon_d = +2t$, while the onsite energies of $p$ sites are $\epsilon_p = 0$, and the small temperature applied is $k_BT = 5\times 10^{-3}t$.  Only nearest-neighbor $d$-$p$ hopping is included.

Figures \ref{fig:checkerboard}(b)-\ref{fig:checkerboard}(d) show our results for a small system with only one primitive cell, where  ED is feasible.  As before, the cluster SP results match well with the ED ones for all three observables studied in this work, while the single-site SP method causes much larger errors. We note that while a larger system is very straightforward to treat with cluster SP simply by using translational symmetry and $k$-point sampling, it is much more difficult to find a good exact benchmark for a large 2D system, especially for the three-band model.  Such a comparison will be a  topic of future work.

\section{Conclusion}
\label{sec:conclusion}
We have introduced a novel non-local slave-particle representation defined on bonds, which improves the slave-particle decomposition by enforcing particle-conserved hoppings on each bond, in addition to the site-centered spinon-slave number matching constraints from previous slave-particle methods.  
We have further developed a cluster approximation for the interacting slave-particle problem based on the expansion and matching of density matrices which maps the slave-particle problem to a set of small overlapping clusters that can be solved self-consistently.  

As a significant improvement to the previous single-site slave-particle theory, our theory correctly predicts the absence of Mott transition in the 1D half-filled Hubbard model (single-site slave-particle methods predict a false  Mott transition).  The method also shows remarkably high accuracy for a wide range of interaction strengths, unit-cell sizes, doping levels, and in both one and two dimensions when compared to exact or high-accuracy benchmark methods.  Computationally, the method is very efficient and requires only a few minutes of CPU time on a serial laptop to find the ground state of the coupled spinon and slave problems.  Future work will benchmark this method more extensively in 2D as well as on more complex multi-orbital real material systems.

\appendix
\section{Mapping and Commutation relations}
\label{app:commute}
Electrons obey the anti-commutation relations
\begin{equation}
    \left\{\hat{c}_{\alpha}, \,\hat{c}_{\alpha^\prime}^\dag\right\} = \delta_{\alpha\alpha^\prime}\,,
    \label{equ:ccommute}
\end{equation}
The spinons ($\hat f_\alpha$) obey the same commutation relations as well, as they are fermions. The slave-particle lowering operator $\hat O$ is then defined to obey the mapping from the original electron Hilbert space to the number-matching states in the enlarged spinon+slave Hilbert space.  The mappings in Eqs.~\eqref{equ:cc2ffOO} and \eqref{equ:c2fO} and~\eqref{equ:cc2ffOO} are made to match all  matrix elements.  That is, for each bond $\avg{\alpha\beta}$, we require
\begin{equation}
\bra{n^\prime} \hat{c}_{\alpha} \ket{n} =
\bra{n^{f\prime}=n^\prime; N^{\prime}=n^\prime} \hat{f}_{\alpha} \hat{O}_{\alpha\avg{\alpha\beta}} \ket{n^f=n; N=n}\,.
\label{equ:c2fOmat}
\end{equation}
Also, the operator $\hat{f}_{\alpha} \hat{O}_{\alpha\avg{\alpha\beta}}$  should still obey  anti-commutation relations like Eq.~\eqref{equ:ccommute}:
\begin{equation}
\begin{split}
\bra{n^{f\prime}=n^\prime; N^{\prime}=&\;n^\prime}
\left\{\hat{f}_{\alpha} \hat{O}_{\alpha\avg{\alpha\beta}}, \hat{f}_{\alpha^\prime}^\dag \hat{O}_{\alpha^\prime\avg{\alpha^\prime\beta^\prime}}^\dag \right\}\\
&\;\times\ket{n^f=n; N=n}
= \delta_{\alpha\alpha^\prime}\,.
\label{equ:fOcommutemat}
\end{split}
\end{equation}
In the following, we are going to prove that the most general form of $\hat{O}_{\alpha\avg{\alpha\beta}}$ shown in Eq.~\eqref{equ:Omat} always obeys the two requirements in Eqs.~\eqref{equ:c2fOmat} and~\eqref{equ:fOcommutemat}.

For the right side of Eq.~\eqref{equ:c2fOmat}, the matrix element can be rewritten as the product of two matrix elements from spinon and slave, separately.
\begin{equation}
\begin{split}
    &\bra{n^{f\prime}=n^\prime; N^{\prime}=n^\prime} \hat{f}_{\alpha} \hat{O}_{\alpha\avg{\alpha\beta}} \ket{n^f=n; N=n} \\
    =\; &\bra{n^{f\prime}=n^\prime} \hat{f}_{\alpha}  \ket{n^f=n}\cdot  \bra{N^{\prime}=n^\prime} \hat{O}_{\alpha\avg{\alpha\beta}} \ket{N=n}
\end{split}
\end{equation}
Now we consider all the cases. When $n_\alpha = 0$, both $\hat{f}_{\alpha}$ and $\hat{c}_{\alpha}$ kill (zero) the matrix element on both sides of Eq.~\eqref{equ:c2fOmat}.  
When $n_\alpha = 1$, since $\hat{O}_{\alpha\avg{\alpha\beta}} \ket{N_{\alpha}=1} = \ket{N_{\alpha}=0}$,  the only non-zero spinon and slave matrix elements come from $n_\alpha^\prime = 0$ and $n_\gamma^\prime = n_\gamma$ for other modes $\gamma \neq \alpha$.  We end up with $\bra{n^{f\prime}=n^\prime} \hat{f}_{\alpha}  \ket{n^f=n}$ for the right side of Eq.  \eqref{equ:c2fOmat}.  Realizing that $\hat{f}_{\alpha}$ and $\hat{c}_{\alpha}$ act identically as same fermionic annihilation operator on identical Hilbert spaces, their matrix elements are the same, so Eq.~ \eqref{equ:c2fOmat} always holds.

For Eq.  \eqref{equ:fOcommutemat}, using the anti-commutation relations of the spinons, we get 
\begin{equation}
\begin{split}
    &\left\{\hat{f}_{\alpha} \hat{O}_{\alpha\avg{\alpha\beta}}, \hat{f}_{\alpha^\prime}^\dag \hat{O}_{\alpha^\prime\avg{\alpha^\prime\beta^\prime}}^\dag \right\} \\
    =\, &\hat{O}_{\alpha^\prime\avg{\alpha^\prime\beta^\prime}}^\dag \hat{O}_{\alpha\avg{\alpha\beta}} \delta_{\alpha\alpha^\prime} + \hat{f}_{\alpha}\hat{f}_{\alpha^\prime}^\dag \left[ \hat{O}_{\alpha\avg{\alpha\beta}},\hat{O}_{\alpha^\prime\avg{\alpha^\prime\beta^\prime}}^\dag \right]
    \label{fOcommute_rewrite}
\end{split}
\end{equation}
The two terms on the right side of Eq.~\eqref{fOcommute_rewrite} are discussed case by case.  

When $\alpha = \alpha^\prime$, using Eq.~ \eqref{equ:Omat} we have 
\begin{equation}
    \hat{O}_{\alpha\avg{\alpha\beta^\prime}}^\dag \hat{O}_{\alpha\avg{\alpha\beta}} = 
    \begin{pmatrix}
        c_{\alpha\avg{\alpha\beta^\prime}}^* c_{\alpha\avg{\alpha\beta}} & 0\\
        0 & 1  
    \end{pmatrix}
    \label{equ:matOO}
\end{equation}
and 
\begin{equation}
    \left[ \hat{O}_{\alpha\avg{\alpha\beta}},\hat{O}_{\alpha\avg{\alpha\beta^\prime}}^\dag \right] = \left(1-c_{\alpha\avg{\alpha\beta^\prime}}^* c_{\alpha\avg{\alpha\beta}}\right)\cdot
    \begin{pmatrix}
        1 & 0\\
        0 & -1  
    \end{pmatrix}
    \,.
    \label{equ:matOOcommute}
\end{equation}
We substitute this into the matrix element on the left side of Eq.~\eqref{equ:fOcommutemat}, and we discuss how the anticommutator acts on the physical ket state $\ket{n^f=n; N=n}$.
If $n_\alpha = 0$ in the physical state, then the action $\hat{f}_{\alpha}\hat{f}_{\alpha}^\dag$ is the identity operation on this state.  The two terms in Eq.~\eqref{fOcommute_rewrite} add to $\hat{O}_{\alpha\avg{\alpha\beta}} \hat{O}_{\alpha\avg{\alpha\beta^\prime}}^\dag$ which acts as identity on for this ket state.  
If $n_\alpha = 1$ in the physical state, the second term of  Eq.~\eqref{fOcommute_rewrite} is zero due to the zeroing action of $\hat{f}_{\alpha}\hat{f}_{\alpha}^\dag$, and the remaining  $\hat{O}_{\alpha\avg{\alpha\beta^\prime}}^\dag \hat{O}_{\alpha\avg{\alpha\beta}}$ acts as identity on this state.  Thus, Eq.~\eqref{equ:fOcommutemat} holds for $\alpha=\alpha'$.

When $\alpha \neq \alpha^\prime$, the first term in Eq.~\eqref{fOcommute_rewrite} is zero.  For the second term, with basis ordered as $\{\ket{N_{\alpha}=0; N_{\alpha^\prime}=0}$, $\ket{N_{\alpha}=0; N_{\alpha^\prime}=1}$, $\ket{N_{\alpha}=1; N_{\alpha^\prime}=0}$, $\ket{N_{\alpha}=1; N_{\alpha^\prime}=1}\}$, we have
\begin{equation}
\begin{split}
    \hat{O}_{\alpha^\prime\avg{\alpha^\prime\beta^\prime}}^\dag \hat{O}_{\alpha\avg{\alpha\beta}} &= 
    \begin{pmatrix}
        0 & 0 & 0 & c_{\alpha^\prime\avg{\alpha^\prime\beta^\prime}}^* \\
        0 & 0 & c_{\alpha^\prime\avg{\alpha^\prime\beta^\prime}}^* c_{\alpha\avg{\alpha\beta}} & 0 \\
        0 & 1 & 0 & 0 \\
        c_{\alpha\avg{\alpha\beta}} & 0 & 0 & 0 
    \end{pmatrix}\\
    &=
    \hat{O}_{\alpha\avg{\alpha\beta}} \hat{O}_{\alpha^\prime\avg{\alpha^\prime\beta^\prime}}^\dag
    \,.
    \label{equ:matOOab}
\end{split}
\end{equation}
Thus the commutator vanishes, Eq.~\eqref{fOcommute_rewrite} is zero in this case, and Eq.~\eqref{equ:fOcommutemat} holds for $\alpha\ne\alpha'$ too.

Based on the above, the commutation relations of the slave-bond operators are 
\begin{equation}
    \left[ \hat{O}_{\alpha\avg{\alpha\beta}},\hat{O}_{\alpha^\prime\avg{\alpha^\prime\beta^\prime}}^\dag \right] = \delta_{\alpha\alpha^\prime} \left(1-c_{\alpha\avg{\alpha\beta^\prime}} c_{\alpha\avg{\alpha\beta}}\right)\cdot
    \begin{pmatrix}
        1 & 0\\
        0 & -1  
    \end{pmatrix}
    \,.
    \label{equ:generalOOcommute}
\end{equation}
This commutation relation means that the ``lowering'' operators $\hat{O}$ are similar to but are \underline{\it not} the same as bosonic field lowering operators.  They are simply defined to obey the mappings of Eq.~\eqref{equ:c2fOmat} and the anticommutation relations of Eq.  \eqref{equ:fOcommutemat}.  Hence, we call them slave-particle (as opposed to slave-boson) operators.

\section{Two-site Hubbard model}
\label{app:2site}
In this appendix, we derive analytical results for the two-site half-filled Hubbard model using three different methods: exact diagonalization of the original fermion Hamiltonian, the single-site slave-particle method, and our cluster slave-bond method.  
\subsection{Exact diagonalization}
\label{app:2site_ED}
The ground state of the half-filled two-site Hubbard model will be a linear combination of the doubly occupied states and the singlet state, so the basis of the subspace is $\ket{\uparrow\downarrow,0}, \ket{0,\uparrow\downarrow}, (\ket{\uparrow,\downarrow}-\ket{\downarrow,\uparrow})/\sqrt{2}$, where the Hamiltonian is represented in this basis as
\begin{equation}
    \hat{H}=
    \begin{pmatrix}
        U & 0 & -\sqrt{2}t\\
        0 & U & -\sqrt{2}t\\
        -\sqrt{2}t & -\sqrt{2}t & 0
    \end{pmatrix}
    \,.
    \label{equ:HED}
\end{equation}
The ground state of this Hamiltonian is 
\begin{equation}
\begin{split}
    \ket{G} =\; &A[4t(\ket{\uparrow\downarrow,0}+ \ket{0,\uparrow\downarrow}) \\
    &+ \left(U+\sqrt{U^2+16t^2}\right)(\ket{\uparrow,\downarrow}-\ket{\downarrow,\uparrow}) ]
    \,,
\end{split}
\end{equation}
where $A$ is the normalization factor.  The corresponding ground-state energy is $E = \left(U-\sqrt{U^2+16t^2}\right)/2$ and the double occupancy on each site is 
\begin{equation}
\begin{split}
D_i~\equiv \avg{\hat{n}_{i\uparrow}\hat{n}_{i\downarrow}} 
= \left(1-\frac{U}{\sqrt{U^2+16t^2}}\right)/4 \,.
\end{split}
\label{equ:dimer_doubleocc}
\end{equation}
Hence the energy is $-2t$ in the non-interacting limit, and becomes $-4t^2/U$ in the large interaction limit.  The double occupancy is $0.25$ and $-2t^2/U^2$ in the same two limits.  So there is not a Mott transition for any finite $U$ based on these exact results.  The analytical results of ground-state energy and double occupancy derived above match with the numerical results in Fig.~\ref{fig:Mott}.

\subsection{Single-site slave-particle}
In the single-site slave-particle method, the total energy is defined as 
\begin{equation}
    E =  -t\sum_{\sigma}\left(\avg{\hat{f}_{1\sigma}^\dag \hat{f}_{2\sigma}} \avg{\hat{O}_{1\sigma}^\dag }\avg{\hat{O}_{2\sigma}} +\text{c.c.}\right) + U \sum_i \avg{\hat{N}_{i\uparrow}\hat{N}_{i\downarrow}}\,.
\end{equation}
To minimize the total energy, the corresponding spinon and the first-site slave-particle Hamiltonian is 
\begin{equation}
\begin{split}
    \hat{H_f} =\; &-t\sum_{\sigma}\left(\avg{\hat{O}_{1\sigma}^\dag }\avg{\hat{O}_{2\sigma}} \hat{f}_{1\sigma}^\dag \hat{f}_{2\sigma} +\text{H.c.}\right) + \sum_{i\sigma} h^\prime_{i\sigma} \hat{N}_{i\sigma}\,,\\
    \hat{H}_{s1} =\;  &-t\sum_{\sigma}\left(\avg{\hat{f}_{1\sigma}^\dag \hat{f}_{2\sigma}} \avg{\hat{O}_{2\sigma}} \hat{O}_{1\sigma}^\dag + \text{H.c.}\right) \\
    &+ U \hat{N}_{1\uparrow}\hat{N}_{1\downarrow} + \sum_\sigma h_{1\sigma} \hat{N}_{1\sigma}\,,
\end{split}
\end{equation}
where $h$ are Lagrange multipliers, while $h^\prime$ combine the Lagrange multipliers and the symmetry-breaking fields to be determined variationally. (The second slave-site Hamiltonian $\hat H_{s2}$ is identical to the first site one other than relabeling.) According to the spin symmetry and inversion symmetry in this system, we conclude that $h^\prime_{i\sigma}$ should be the same for all sites and spins, and the same statement also holds for $h_{i\sigma}$.  Based on this symmetry analysis, the spinon Hamiltonian is solved easily by diagonalizing $2\times2$ matrices for each spin channel.  As a result, the occupancy is $\avg{n_{i\sigma}} = 1/2$ for every site and spin, and the spinon density matrix element of interest is $\avg{\hat{f}_{1\sigma}^\dag \hat{f}_{2\sigma}} = 1/2$.  

Next, in the basis $\ket{0}, \ket{\uparrow}, \ket{\downarrow}, \ket{\uparrow\downarrow}$, we have 
\begin{equation}
    \hat{H}_{s1}=
    \begin{pmatrix}
        0 & -y & -y & 0\\
        -y & h & 0 & -y\\
        -y & 0 & h & -y\\
        0 & -y & -y & 2h+U
    \end{pmatrix}
    \,,
\end{equation}
where we set $h_{i\sigma}=h$ based on the symmetry discussed above, and $y\equiv 0.5t(1+c)\avg{O_{2\sigma}}$, in which $c$ is the real $c$-gauge number determined to make $\avg{O_{i\sigma}}=1$ in the non-interacting limit.  

When $U=0$, the ground state of the slave-particle Hamiltonian is 
$$
\ket{G_0} = \left(\ket{0}+ \ket{\uparrow}+ \ket{\downarrow}+ \ket{\uparrow\downarrow}\right)/2 \,.
$$
In order to have the spinon hopping renormalization factor $\avg{O_{i\sigma}}=1$ with this state, the $c$-gauge number is $c=1$, and thus $y=t\avg{O_{2\sigma}}$.  

For a finite interaction $U/t>0$, to obey the particle number matching constraint in Eq.~\eqref{equ:constraint}, we find that $h = -U/2$.  So the ground state with finite interaction is 
$$
\ket{G} = (\left\ket{0}+ a\ket{\uparrow}+ a\ket{\downarrow}+ \ket{\uparrow\downarrow}\right)/\sqrt{2(1+a^2)} \,.
$$
where $a\equiv \sqrt{(8t+U)/(8t-U)}$ when $U<8t$, while $a\rightarrow\infty$ when $U\geq8t$.  Finally, using the single-site slave-particle method, when $U<8t$, the ground-state energy is $E = -(U-8t)^2/32t$, and the double occupancy is $D_i = (8t-U)/32t$, while when $U\geq 8t$, both ground-state energy and double occupancy are zero, which match with the numerical results in Fig.~\ref{fig:Mott}.  This result clearly indicates a false Mott transition at $U=8t$ in contrast to the exact diagonalization method.  

\subsection{Cluster slave-particle method}
In the cluster slave-particle method, the total energy is defined as 
\begin{equation}
    E =  -t\sum_{\sigma}\left(\avg{\hat{f}_{1\sigma}^\dag \hat{f}_{2\sigma}} \avg{\hat{O}_{1\sigma}^\dag \hat{O}_{2\sigma}} +\text{c.c.}\right) + U \sum_i \avg{\hat{N}_{i\uparrow}\hat{N}_{i\downarrow}}\,,
    \label{equ:dimer_CSP_E}
\end{equation}
where we hide the bond index for $\hat{O}$ operators because there is only one bond, and the $c$ gauge for spin up and spin down are the same by symmetry in this system.  
To minimize the total energy, using a two-site cluster, the corresponding spinon and the cluster slave-particle Hamiltonian are 
\begin{equation}
\begin{split}
    \hat{H_f} =\; &-t\sum_{\sigma}\left(\avg{\hat{O}_{1\sigma}^\dag \hat{O}_{2\sigma}} \hat{f}_{1\sigma}^\dag \hat{f}_{2\sigma} +\text{H.c.}\right) + \sum_{i\sigma} h^\prime_{i\sigma} \hat{N}_{i\sigma}\,,\\
    \hat{H}_{C} =\;  &-t\sum_{\sigma}\left(\avg{\hat{f}_{1\sigma}^\dag \hat{f}_{2\sigma}}  \hat{O}_{1\sigma}^\dag \hat{O}_{2\sigma} + \text{H.c.}\right) \\
    &+ \sum_i \left(U \hat{N}_{i\uparrow}\hat{N}_{i\downarrow} + \sum_\sigma h_{i\sigma} \hat{N}_{i\sigma}\right)\,.
\end{split}
\end{equation}
Similarly, $h$ are Lagrange multipliers, while $h^\prime$ combine the Lagrange multipliers and the symmetry-breaking fields to be determined variationally.  The spin symmetry and inversion symmetry also hold here, and the spinon Hamiltonian can be solved easily as before: we find  $\avg{n_{i\sigma}} = 1/2$ and $\avg{\hat{f}_{1\sigma}^\dag \hat{f}_{2\sigma}} = 1/2$.  

The non-interacting slave Hamiltonian can be separated by spin channels, where the spin $\sigma$ Hamiltonian is represented in the basis ordered as $\ket{0, 0}, \ket{\sigma, 0}, \ket{0, \sigma}, \ket{\sigma, \sigma}$
\begin{equation}
    \hat{H}_{C\sigma}=
    \begin{pmatrix}
        0 & 0 & 0 & 0\\
        0 & h & -y & 0\\
        0 & -y & h & 0\\
        0 & 0 & 0 & 2h
    \end{pmatrix}
    \,,
\end{equation}
where $y = 0.5t(1-C^2)$, and $C$ stands for the gauge to be determined and is assumed to be real.  We find that, in order to have the non-interacting spinon hopping renormalization factor $\avg{\hat{O}_{1\sigma}^\dag \hat{O}_{2\sigma}} = 1$ and the particle number match with the spinon result, the $c$-gauge number must be $C = \sqrt{3}$.  

With a finite interaction $U$, matching particle numbers give the final slave-particle ground state  
\begin{equation}
\begin{split}
    \ket{G} =\; &A[-4t(\ket{\uparrow\downarrow,0}+ \ket{0,\uparrow\downarrow}) \\
    &+ \left(U+\sqrt{U^2+16t^2}\right)(\ket{\uparrow,\downarrow}+\ket{\downarrow,\uparrow}) ]
    \,,
\end{split}  
\label{equ:dimer_CSP_GS}
\end{equation}
where $A$ is the normalization factor.  The corresponding renormalization factor is $\avg{\hat{O}_{1\sigma}^\dag \hat{O}_{2\sigma}} = 4t/\sqrt{U^2+16t^2}$.  In other words, the inter-site hopping (and thus renormalization factors) of the slave-bond is always non-zero in this system which indicates an absence of a Mott transition.  Finally, observables can be calculated using the slave-particles ground-state expression \eqref{equ:dimer_CSP_GS}.  The ground-state energy given by Eq.~\eqref{equ:dimer_CSP_E} is $E = \left(U-\sqrt{U^2+16t^2}\right)/2$.  The double occupancy on each site is 
\begin{equation}
    D_i~\equiv \avg{\hat{N}_{i\uparrow}\hat{N}_{i\downarrow}} 
    = \left(1-\frac{U}{\sqrt{U^2+16t^2}}\right)/4 \,,
\end{equation}
where the double occupancy is always non-zero for arbitrary $U/t$.  Finally, note that the expressions of ground-state energy and double occupancy are exactly the same as the exact diagonalization results derived above in Appendix~\ref{app:2site_ED}.

\section{Formalism for $c$-gauge}
\label{app:cgauge}
In this appendix, we derive the analytical formulas of Eqs.~\eqref{equ:cintra} and \eqref{equ:cinter} for the $c$-gauge numbers.  As we introduced in Sec.~\ref{sec:decomposition}, the $c$ gauges are chosen (i) to forbid all unphysical intracluster hoppings in slave problem that change particle numbers, and (ii) to ensure that the spinon problem recovers the original fermion problem in the non-interacting limit.  

\subsection{Intracluster hoppings}
\begin{figure}[t!]
\begin{center}
\includegraphics[scale=0.8]{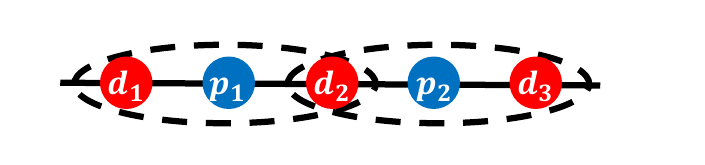}
\end{center}
\caption{
 An illustration of the one-dimensional $d$-$p$ chain, where red atoms represent $d$ sites while blue atoms represent $p$ sites, and each atom is labeled by white letters.   The black dashed ellipses are the clusters used in the cluster slave-bond calculation.
}
\label{fig:supp1}
\end{figure}

By definition, both sites of an intracluster hopping can be found in one same cluster.  Note that even though the nearest-neighbor hopping such as the $d_2$-$p_2$ hopping in Fig.~\ref{fig:supp1} can appear as an outward hopping with respect to the cluster $[d_1p_1d_2]$, it is still treated as a short-ranged intracluster hopping, whose $c$ gauge is computed in cluster $[d_2p_2d_3]$.
For intracluster hoppings, the first constraint (i) requires $c_{\alpha\avg{\alpha\beta}} + c_{\beta\avg{\alpha\beta}}=0$, and the second (ii) requires $\avg{\hat{O}^\dag_{\alpha \avg{\alpha\beta}} \hat{O}_{\beta \avg{\alpha\beta}}}=1$ in the non-interacting calculation.  For each bond $\avg{\alpha\beta}$, there are two $c$-gauge variables to be determined, obeying the above two equations.  By solving the two equations, we find 
\begin{equation}
    c_{\alpha\avg{\alpha\beta}} = \sqrt{1-\frac{1}{\avg{0,1|\hat\rho_{\alpha\beta}^0|1,0}}}\,,
    \label{equ:cintra}
\end{equation}
and $\hat\rho_{\alpha\beta}^0$ is a non-interacting bond density matrix, where the 0 and 1 in the bras and kets are the occupation numbers on the slave modes $\alpha$ and $\beta$, respectively.  The superscript 0 represents the operators in non-interacting slave-particle calculations.  The bond density matrix is defined by tracing out other degrees of freedom in the cluster density matrix 
$
    \hat\rho_{\alpha\beta}^0 \equiv \Tr_{\gamma\in\mathcal{C}|\gamma\neq\alpha,\beta}\left(\hat\rho_{\mathcal{C}}^0\right)\,.
$

\subsection{Intercluster hoppings}
Aside from the intracluster hoppings, the other hoppings are long-ranged intercluster hoppings, such as the $p_1$-$p_2$ hopping in Fig.~\ref{fig:supp1}.  Based on the density matrix expansion in Eq.~\eqref{equ:expan}, an intercluster hopping $\avg{\alpha\beta}$ in our slave-particle approach is decoupled as $\avg{\hat{O}_{\alpha \avg{\alpha\beta}}^\dag \hat{O}_{\beta \avg{\alpha\beta}}} = \avg{\hat{O}_{\alpha \avg{\alpha\beta}}^\dag} \avg{\hat{O}_{\beta \avg{\alpha\beta}}}$.  The constraints on the bond $\avg{\alpha\beta}$ are $\avg{\hat{O}_{\alpha \avg{\alpha\beta}}}_{\rho^0}=\avg{\hat{O}_{\beta \avg{\alpha\beta}}}_{\rho^0}=1$ in order to recover the non-interacting Hamiltonian with the spinons alone ($\rho^0$ is the non-interacting slave density matrix).   Each of these constraints is easily solved to get
\begin{equation}
    c_{\alpha\avg{\alpha\beta}} = \frac{1}{\avg{0|\hat\rho_{\alpha}^0|1}}-1\,,
    \label{equ:cinter}
\end{equation}
where 0 and 1 are the occupation numbers of the slave mode $\alpha$.  The single-orbital density matrix is determined as 
\begin{equation}
    \hat\rho_{\alpha}^0\equiv \frac{1}{M_\alpha}\sum_{\mathcal{C}|\alpha\in\mathcal{C}}\Tr_{\gamma\in\mathcal{C}|\gamma\neq \alpha}(\hat\rho_\mathcal{C})\,,
    \label{equ:rhoi_cluster}
\end{equation}
where $M_\alpha$ is the number of clusters overlapping on the slave mode $\alpha$.

\begin{widetext}
\section{Redundancy of single-site density matrix constraints}
\label{app:MoreConstraints}
In this appendix, we describe in detail the constraints ensuring that the single-site density matrix $\hat\rho_i$ at a shared site $i$ among multiple clusters is described consistently among them.  As explained in the main text, these constraints are imposed by the Lagrange multipliers $\hat{\Lambda}_i^{(\mathcal{C})}$ and associated energy terms in the function $F$ defined in Eq.~\eqref{equ:clusterF} that are minimized.  The key result of this appendix is that all these Lagrange multipliers $\hat{\Lambda}_i^{(\mathcal{C})}$ can be set to zero. 

Firstly, there are some redundant Lagrange multipliers in the above set of constraints. For a site $i$, there are $M_i$ Lagrange multiplier matrices $\hat{\Lambda}_i^{(\mathcal{C})}$ since there are $M_i$  distinct clusters overlapping on the site $i$.  The task of the $\hat{\Lambda}_i^{(\mathcal{C})}$ is to make all the site density matrices equal to each other (and equal to $\hat\rho_i$),  $\hat\rho^{(\mathcal{C})}_i=\hat\rho_i$.  This means that we only require $M_i-1$ distinct Lagrange multiplier matrices at site $i$, and we can pick one of the clusters to be a reference cluster $\mathcal{C}^*_i$ whose single-site density matrix the other clusters must match.  For example, if a site $i$ is not shared by multiple clusters (so $M_i=1$), then no Lagrange multipliers should be added to match the single-site density matrix to itself; if $M_i=2$, then we need only match the density matrix of the second cluster with the first. We can safely set the Lagrange multiplier matrix for our reference cluster to zero, 
 $\hat{\Lambda}_i^{(\mathcal{C}^*_i)} = 0$.

Secondly, the density matrix matching constraints are obeyed throughout the entire minimization process.  This means that not only $\hat\rho_i^{(\mathcal{C})}=\hat\rho_i$ at the beginning of the minimization, but also that $\delta\hat\rho_i^{(\mathcal{C})}=\delta\hat\rho_i$ when minimizing the function $F$ along the gradient.  The gradient of $F$ along $\hat\rho_{\mathcal{C}}$ is 
\begin{equation}
\begin{split}
    \frac{\partial F}{\partial \hat\rho_{\mathcal{C}}} =
    &\;-\sum_{\alpha\beta | \avg{\alpha\beta}\in\mathcal{C}} t_{\alpha\beta} \avg{\hat{f}^\dag_{\alpha} \hat{f}_{\beta}}_{\rho_f} \hat{O}^\dag_{\alpha\avg{\alpha\beta}} \hat{O}_{\beta\avg{\alpha\beta}}
    -\sum_{\substack{\beta | \beta\in\mathcal{C} \\ \alpha | \avg{\alpha\beta}\notin\forall\mathcal{C}^\prime}} \frac{t_{\alpha\beta}}{M_\beta} \avg{\hat{f}^\dag_{\alpha} \hat{f}_{\beta}}_{\rho_f}  \left[\avg{\hat{O}^\dag_{\alpha\avg{\alpha\beta}}}_{\rho_{i|\alpha\in i}} \hat{O}_{\beta\avg{\alpha\beta}}+\text{H.c.}\right]\\
    &\;+\sum_{i|i\in\mathcal{C}} \frac{1}{M_i} \hat{H}^{\text{int}}_i
    +\sum_{\alpha|\alpha\in\mathcal{C}} h_{\alpha}^{(\mathcal{C})} \hat{N}_{\alpha}-\epsilon _{\mathcal{C}}\hat I_{\mathcal C}
    +\sum_{i| {i}\in\mathcal{C}} (1-1/M_i)\hat{\Lambda}_i^{(\mathcal{C})} 
        \,.
\end{split}
\label{equ:pF_prhoC}
\end{equation}
We collect the derivatives versus all $N$ clusters into the gradient $\nabla F = (\partial F/\partial\hat\rho_{\mathcal{C}_1},\partial F/\partial\hat\rho_{\mathcal{C}_2},\ldots,\partial F/\partial\hat\rho_{\mathcal{C}_N})$.  As is standard in the Lagrange multiplier method, we will project out the part of $\nabla F$ that breaks the constraints to determine the value of the Lagrange multipliers.   This is most easily done by considering a step of size $\eta$ along the gradient $\left( \delta\hat\rho_{\mathcal{C}_1}, \delta\hat\rho_{\mathcal{C}_2},...,\delta\hat\rho_{\mathcal{C}_N} \right) = \eta \nabla F $, so the change $\delta\hat\rho^{(\mathcal{C})}_i$ is given by
\begin{equation}
\delta\hat\rho_i^{(\mathcal{C})}
    = \Tr_{\mathcal{C}-i}(\delta\hat\rho_{\mathcal{C}}) 
    = \eta \Tr_{\mathcal{C}-i}\left( \frac{\partial F}{\partial \hat\rho_{\mathcal{C}}}\right)\,,
\end{equation}
where, again, $\Tr_{\mathcal{C}-i}$ denotes a trace over the cluster $\mathcal{C}$ but excluding the degrees of freedom on site $i$.    The key enabling observation allowing us to move forwards is that the hopping operators $\hat O,\hat O^\dagger$ are traceless.  Hence, the intracluster hopping terms in Eq.~\eqref{equ:pF_prhoC} involve hoppings between different sites so they have zero traces, the intercluster hopping terms are zero unless the state $\beta$ is on the site $i$, and the remaining terms are local single-site operators.  At this stage, we have
\begin{equation}
\begin{split}
    \delta\hat\rho_i^{(\mathcal{C})}
    = 
    &\;\eta\left( -\sum_{\substack{\beta | \beta\in i \\ \alpha | \avg{\alpha\beta}\notin\forall\mathcal{C}^\prime}} \frac{t_{\alpha\beta}}{M_\beta} \avg{\hat{f}^\dag_{\alpha} \hat{f}_{\beta}}_{\rho_f}  \left[\avg{\hat{O}^\dag_{\alpha\avg{\alpha\beta}}}_{\rho_{k|\alpha\in k}} \hat{O}_{\beta\avg{\alpha\beta}}+\text{H.c.}\right] + \frac{\hat{H}^{\text{int}}_i}{M_i}\right. \\
    &\;\left. +\sum_{\alpha|\alpha\in i} h_{\alpha}^{(\mathcal{C})} \hat{N}_{\alpha}-\epsilon _{\mathcal{C}}\hat I_i
    + (1-1/M_i)\hat{\Lambda}_i^{(\mathcal{C})} 
     + q_i^{(\mathcal{C})}\hat I_i
    \right)
     \prod_{j\in\mathcal{C}|j\neq i} S_j\,,
\end{split}
\label{equ:delta_rhoiC}
\end{equation}
where $q^{(\mathcal{C})}_i$ is a potentially cluster-dependent number coming from traces of local operators not on site $i$, and $S_j$ is the dimension of the single-site density matrix $S_j \equiv \Tr_j(\hat{I}_j)$ (i.e., dimension of the Hilbert space on site $j$).  A moment's reflection shows that the first two terms in Eq.~\eqref{equ:delta_rhoiC} are independent of the cluster $\mathcal{C}$, and we combine them into an operator $\hat X_i$.  Upon creating new constants $r_i^{(\mathcal{C})}=\epsilon_C+q^{(\mathcal{C})}_i$ and $\zeta=\eta\prod_{j\in\mathcal{C}|j\neq i}
S_j$, we  have 
\begin{equation}
    \delta\hat\rho_i^{(\mathcal{C})}    = 
    \zeta\left( \hat X_i
     +\sum_{\alpha|\alpha\in i} h_{\alpha}^{(\mathcal{C})} \hat{N}_{\alpha}
    + (1-1/M_i)\hat{\Lambda}_i^{(\mathcal{C})} 
     + r_i^{(\mathcal{C})}\hat I_i
    \right)
     \label{equ:delta_rhoiCclean}
\end{equation}
As per Eq.~\eqref{equ:rhoiform}, we sum this $\delta\hat\rho_i^{(\mathcal{C})}$ over all cluster $\mathcal{C}$ containing site $i$ and divide by $M_i$ to find an analogous expression for $\delta\hat\rho_i$
\begin{equation}
    \delta\hat\rho_i = 
    \zeta\left( \hat X_i
     +\sum_{\alpha|\alpha\in i} \bar h_{\alpha} \hat{N}_{\alpha}
    + (1-1/M_i)\overline{\hat\Lambda_i }
     + \bar r_i\hat I_i
    \right)
     \label{equ:delta_rhoi}
\end{equation}
where overbars mean averaging over the clusters.  Equating Eqs.~\eqref{equ:delta_rhoiCclean} and \eqref{equ:delta_rhoi} yields
\begin{equation}
    \hat\Lambda_i^{(\mathcal{C})} = 
    \overline{\hat\Lambda_i} +
    (1-1/M_i)^{-1}\sum_{\alpha|\alpha\in i} (\bar h_\alpha-h_{\alpha}^{(\mathcal{C})}) \hat{N}_{\alpha} + (1-1/M_i)^{-1}(\bar r_i - r_i^{(\mathcal{C})})\hat I_i\,.
\end{equation}
The second and third terms above are either redundant or irrelevant: the second term involves shifts of onsite state energies on site $i$ which can be absorbed into the $h^{(\mathcal{C})_\alpha}$ Lagrange multipliers that enforce spinon and slave-particle number matching, while the third term represents a shift of the cluster Hamiltonian $\hat H_\mathcal{C}$ by a constant which does not change its eigenfunctions, shift all eigenenergies by the same amount, and thus does not change the thermal density matrix $\hat\rho_\mathcal{C}$ computed from the Boltzmann distribution.  Thus, the only remaining meaningful term is $\hat\Lambda_i^{(\mathcal{C})} = \overline{\hat\Lambda_i}$ which is cluster independent: since we have a reference cluster for which $\Lambda_i^{\mathcal{C}_i^*}=0$, all the $\hat\Lambda_i^{(\mathcal{C})}=0$. 

Hence, we conclude that the mean particle-number-matching constraints between spinon and slave particles are sufficient for describing overlapping clusters.  Additional constraints, beyond the automatic mean particle-number matching, are redundant.

\end{widetext}

\section{Tests on single-band Hubbard models}
\label{app:testHubbard}

\begin{figure}[t]
\begin{center}
\includegraphics[scale=0.35]{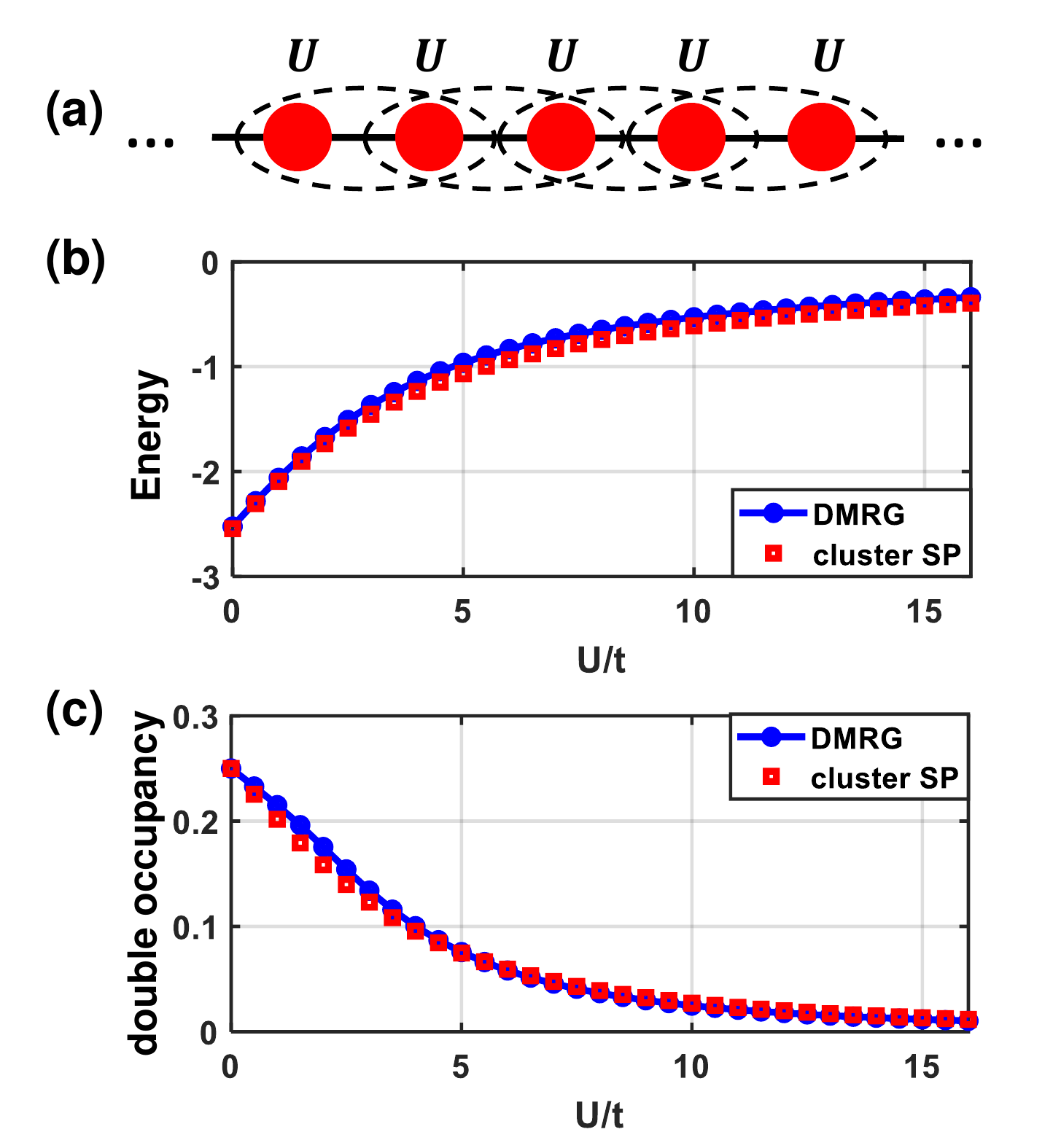}
\end{center}
\caption{
(a) Illustration of the one-dimensional Hubbard model with periodic boundary conditions.   The black dashed ellipses are the clusters used in the cluster slave-particle calculation.  
(b), (c) The total energy per two-site unit cell in units of $t$, and the double occupancy versus the interaction strength $U$, respectively.  
All results are for a 64-site 1D single-band Hubbard system at half filling: blue circles represent DMRG results with OBC, while red squares represent cluster SP results assuming a translational period of two sites (PBC).
}
\label{fig:1DHubbard}
\end{figure}

In this appendix, we show our benchmarking results for 1D and 2D half-filled single-band Hubbard models, i.e., when there is one correlated orbital at each site.  This type of model is a ``$d$''-only model when describing a transition metal oxide since the oxygen orbitals are removed from the model.  {\it A priori} we expect this type of model to be less accurate than the $d$-$p$ models studied in the main text for at least two reasons: (i) the removal of degrees of freedom can be accommodated by renormalizing the remaining parameters of the model (e.g., nearest neighbor hopping element $t$), but this type of process is always approximate, and (ii) removal of the oxygen $p$ orbitals gives rise to effective orbitals (e.g., Wannier states) on the $d$ sites that are much less localized than the starting orbitals of the $d$-$p$ model, so that using an on-site only form of the interaction is much less justified for the $d$-only case than for the $d$-$p$ case (i.e., the form of the Hamiltonian is less accurate in the $d$ only case with only on-site interactions).  However, the $d$-only model is a standard benchmark in the field so comparisons to it are still helpful for connecting to prior literature.

As is standard for one-band Hubbard models \cite{essler2005one}, the nearest neighbor hopping strength is set to be $-t$, while all other hoppings are set to zero.  The onsite energy of each $d$ site is set to zero, and the local on-site interaction strength is set to $U$.   In addition, a small temperature of $k_BT = 5\times 10^{-3}t$ is introduced in the cluster slave-particle calculations in an identical manner to that for the $d$-$p$ models in Sec.~\ref{sec:test1d}.  (From a practical viewpoint, to simulate the $d$-only systems without having to modify our $d$-$p$ software implementation, we choose a $d$-$p$ model where there are no hoppings involving the $p$ orbitals which have a very low energy and are filled with electrons and thus are completely inert, while direct $d$-$d$ hoppings $-t$ are introduced for the remaining electrons in the $d$ manifold.)

The first set of tests is on a 1D 64-site single-band Hubbard model with periodic boundary conditions, whose lattice is illustrated in Fig.~\ref{fig:1DHubbard}(a).  The clusters used in the cluster SP calculation are marked by black dashed ellipses and consist of two neighboring $d$ sites and neighboring clusters overlap with each other.  For cluster SP calculations with PBC, we assume a translational period of two sites (i.e., a two-site unit cell), while for DMRG benchmarks, we use OBC as discussed in Sec.\ref{sec:systemsize}.  Each data point requires only about 10 CPU seconds for the cluster slave-particle calculations but needs about 7 hours for the DMRG calculations. 

Although the two-site SP clusters in these tests are smaller than the three-site SP clusters used in $d$-$p$ model testing in Sec.~\ref{sec:test1d}, the two-site clusters are capable of reproducing high-quality results as shown in Fig.~\ref{fig:1DHubbard}.  Noticing that the double occupancy of both DMRG and cluster SP are finite regardless of interaction strength, we correctly predict the absence of a Mott transition in this 1D model  in contrast to the false Mott transition predicted in single-site slave-particle theory and DMFT \cite{georgescu2017symmetry, lee2019rotationally, bolech2003cellular, zgid2012truncated}.

\begin{figure}[t]
\begin{center}
\includegraphics[scale=0.33]{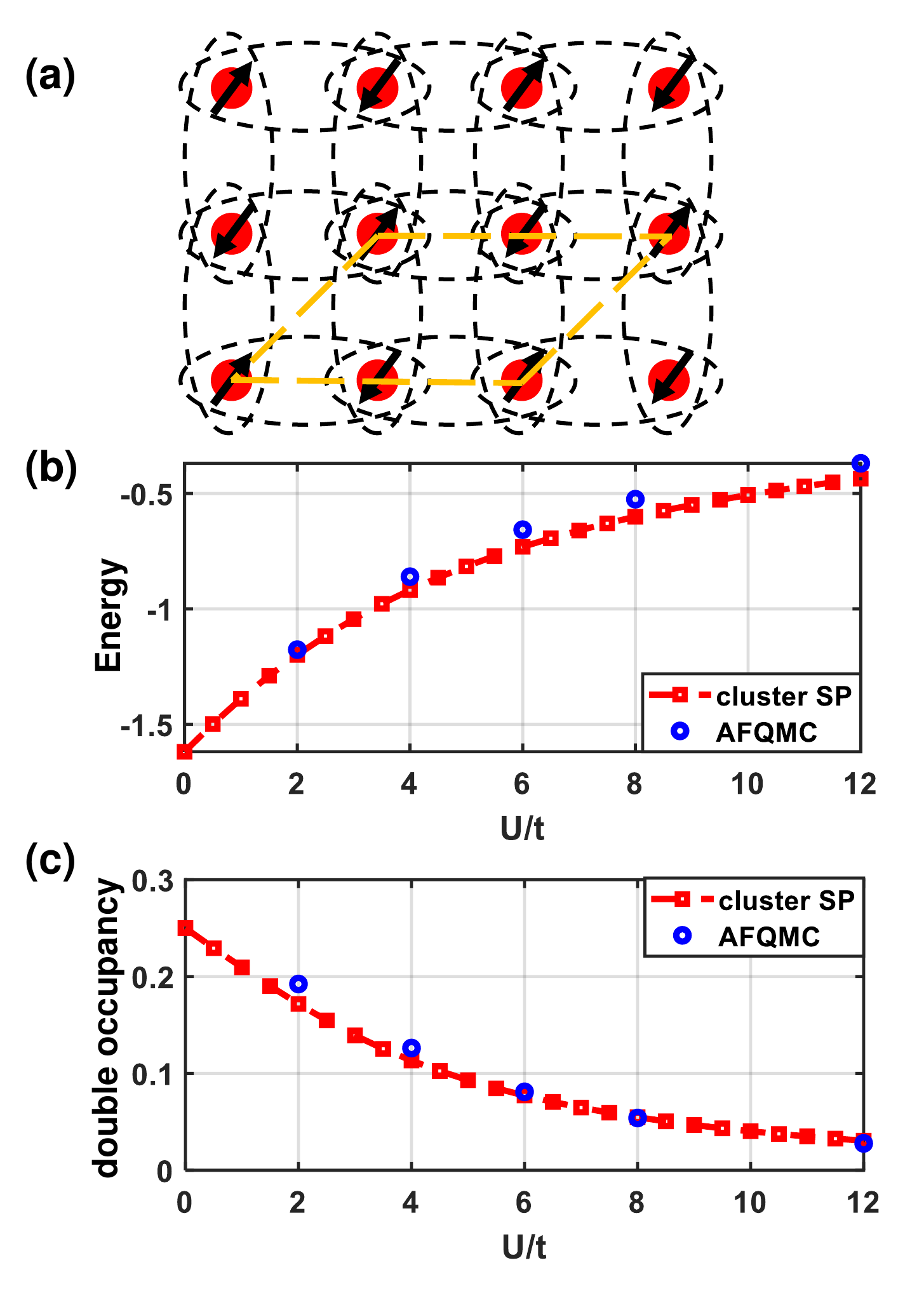}
\end{center}
\caption{
(a) Illustration of the two-dimensional Hubbard model with periodic boundary conditions.   The black dashed ellipses are the clusters used in the cluster slave-particle calculation, and the orange dashed parallelogram represents the two-site primitive cell under the N\'eel AFM correlation.
(b), (c) The total energy per two-site unit cell in units of $t$, and the double occupancy versus the interaction strength $U$, respectively.  
The red squares represent cluster SP results generated by a ($64\times 64$)-site system at half-filling with PBC, while the blue circles are auxiliary-field quantum
Monte Carlo (AFQMC) benchmarks \cite{leblanc2015solutions} in the thermodynamic limit. 
}
\label{fig:2DHubbard}
\end{figure}

Next, we examine a two-dimensional $64\times 64$-site one-band Hubbard model at half filling with PBC whose lattice is illustrated in Fig.~\ref{fig:2DHubbard}(a).  The two-site clusters marked by black dashed ellipses and the two-site N\'eel-ordered unit cell marked by the orange dashed parallelogram are chosen in the same manner as for the 2D $d$-$p$ model of Sec.~\ref{sec:2D}.  Figures \ref{fig:2DHubbard}(b) and \ref{fig:2DHubbard}(c) show the total energy and double occupancy results of the cluster SP method.  As a comparison, we also show thermodynamic limit results from auxiliary-field quantum Monte Carlo (AFQMC) calculations \cite{leblanc2015solutions}.  Using SP clusters containing only two sites, the cluster SP method shows remarkable quantitative accuracy with a modest computational cost of only about two serial CPU minutes for each data point.

\begin{figure}[t]
\begin{center}
\includegraphics[scale=0.35]{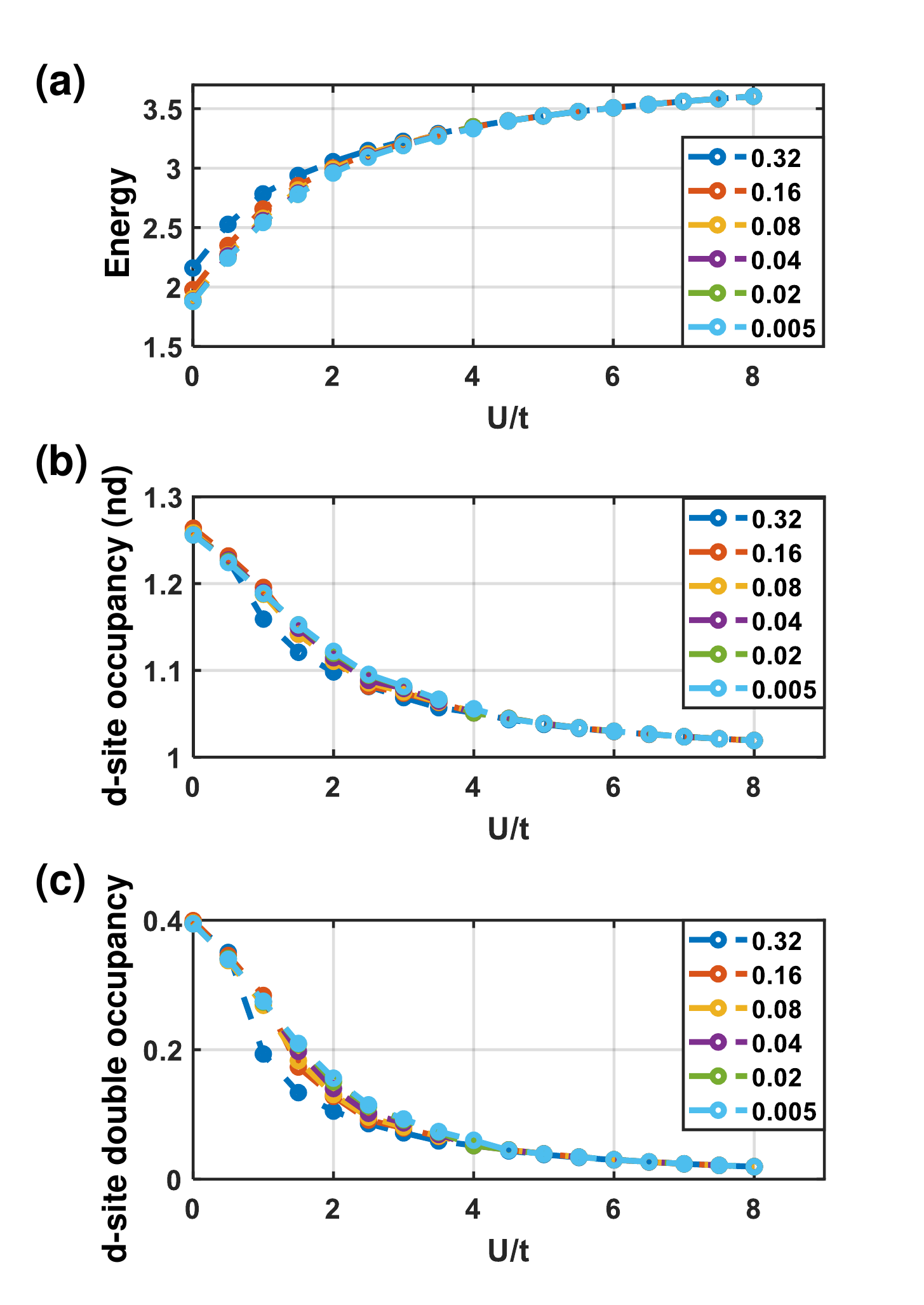}
\end{center}
\caption{
 (a) Total energy per unit cell (units of $t$), (b) $d$-site occupancy, and (c) $d$-site double occupancy versus the interaction strength $U$, respectively, for a 64-site 1D $d$-$p$ system with cluster SP method and assuming translational unit cell of four sites.  Different curves show results with different temperatures marked by the legend, where the temperatures $k_BT$ are in units of $t$.
}
\label{fig:temperature}
\end{figure}

\section{Finite-temperature error}
\label{app:temperature}

Generally, the partial trace of a pure state will be a mixed state.  Thus, even if our current cluster SP method only aims to predict the ground-state (pure-state) properties of the entire (lattice) system, in practice we use a finite temperature in all the slave-particle cluster calculations.  This is because the mixed state introduced by the Boltzmann distribution can better capture the cluster's statistical properties.  

The use of a small but finite-temperature Boltzmann distribution is a straightforward way to generate a mixed state made from a number of low-energy eigenstates of the cluster Hamiltonian, but the finite temperature itself introduces errors due to the inclusion of higher-energy excited states.  Here we show that the errors are quite small in magnitude and easily controllable.

Numerically, the error in the total energy per unit cell has the same order of magnitude as the temperature itself as shown in Fig.~\ref{fig:temperature}(a).  For example, compared to the benchmark $T=0$ DMRG results in Fig.~\ref{fig:64site}, the largest error caused by $k_BT=0.32t$ is $0.28t$, while the largest error caused by $k_BT=0.16t$ is $0.10t$.  For this system, the finite temperature error in energy gets almost negligible when $k_BT$ is below $0.04t$.  Similarly, the $d$-site occupancy and double occupancy are quite accurate at $k_BT=0.04t$, as shown in Figs.~\ref{fig:temperature}(b) and \ref{fig:temperature}(c).  All results in the main text are based on an even lower temperature of $k_BT=0.005t$, so it is safe to ignore the finite-temperature effect in our results.

From a practical viewpoint, starting from an arbitrary initial setup, a high-temperature calculation typically converges much faster than a low-temperature calculation.  Hence, for any given system and set of parameters, we begin with a high-temperature calculation and use its self-consistent converged solution as the starting point for a lower-temperature calculation (and repeat the process for ever lower temperatures), a process that leads to significant computational efficiency.

\begin{figure}[t]
\begin{center}
\includegraphics[scale=1]{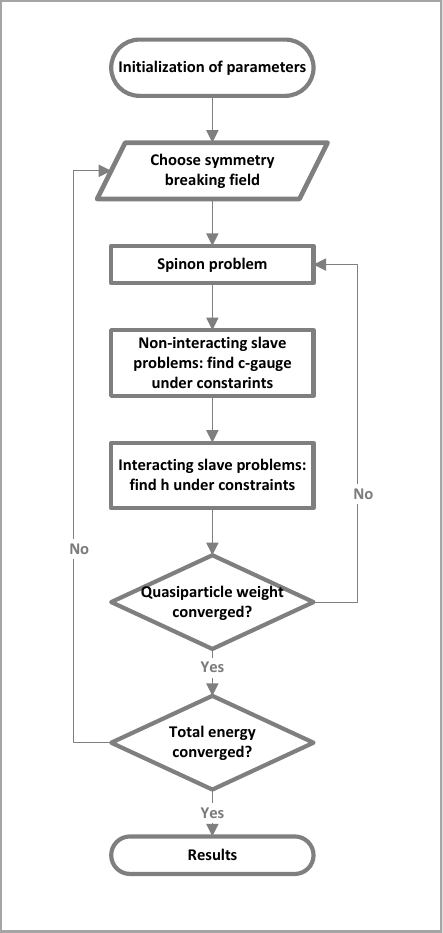}
\end{center}
\caption{
The workflow of a typical calculation based on the slave-particle approach.  
}
\label{fig:workflow}
\end{figure}
\section{Workflow}
\label{app:workflow}

In this section, we describe the workflow of a typical slave-particle calculation in Fig.~\ref{fig:workflow}.  
It contains the following steps:
\begin{enumerate}
    \item Initialize all parameters, including all the Lagrange multipliers $h$, $c$-gauge numbers, hopping renormalization factors $\avg{\hat{O}^\dag_{\alpha\avg{\alpha\beta}} \hat{O}_{\beta\avg{\alpha\beta}}}$ and $\avg{\hat{O}_{\alpha\avg{\alpha\beta}}}$.  When starting from scratch, reasonable {\it a priori} choices are zero for $h$, and unity for $c$ and the renormalization factors.
    
    \item For the first iteration, guess some symmetry-breaking field, typically small random numbers much smaller in magnitude than $t$ or $U$.  For the following iterations, choose the symmetry-breaking field variationally by gradient descent of the total energy.  The updating of the symmetry-breaking fields is the major outer loop of energy minimization.
    \label{item:bigB}
    \item Solve the spinon Hamiltonian in Eq.~\eqref{equ:Hfc}, then compute the hopping renormalization factors $\avg{\hat{f}^\dag_{\alpha} \hat{f}_{\beta}}$ for use in the slave problem in the next step and the spinon occupation numbers $\avg{\hat{n}_{\alpha}}$.
    \label{item:spinon}
    \item Solve the non-interacting slave Hamiltonian in Eq.~\eqref{equ:Hfc}, and search for $c$-gauge number to obey the corresponding constraints.  This is most easily done via Newton's method.
    \item Solve the interacting slave Hamiltonian and search for the Lagrange multipliers $h$ under the corresponding constraints.  Note that these $h$ will be generally different from the non-interacting $h$ in the previous step, although the non-interacting ones provide a good initial guess when starting from scratch.
    
    \item If the renormalization factors $\avg{\hat{O}^\dag_{\alpha\avg{\alpha\beta}} \hat{O}_{\beta\avg{\alpha\beta}}}$ and $\avg{\hat{O}_{\alpha\avg{\alpha\beta}}}$ differ from the prior step by more than a small convergence tolerance, go back to step~\ref{item:spinon} while using the updated renormalization factors, and repeat the calculation until tolerance is achieved.  
    
    \item If the total energy is not converged with respect to the symmetry-breaking field (i.e., the successive change of energy is too large between updates of symmetry-breaking fields), then return to step~\ref{item:bigB} and update the symmetry-breaking field to minimize the total energy.  
\end{enumerate}

We note that methods such as Newton's method or gradient descent require derivatives, and since they are computed numerically by finite differences (absent analytical expressions for derivatives at present time), in practice we encapsulate inner loops (e.g., the non-interacting slave solver) as routines that are called repeatedly to compute numerical finite-difference derivatives.

Based on the workflow chart above, we can estimate the computational cost needed to complete the full calculation.  First, the single-particle spinon Hamiltonian is practically very simple and a variety of methods can be unleashed to find its ground state (direct diagonalization is simplest but cubic in system size while linear-scaling methods are available if needed in the very large system limit~\cite{goedecker_linear_1999}).
For the more complex cluster slave-particle problem, the computational cost is $\propto S_C N_C N_L$, where $S_C$ is the computational cost of the exact diagonalization of one cluster, $N_C$ is the number of clusters in one unit cell, and $N_L$ is the total number of cluster solutions needed (a loop count).  $N_L$ is the product of the number of Lagrange multipliers $h$, the number of self-consistent steps needed to converge the renormalization factors, and the number of gradient descent steps needed to converge the symmetry-breaking fields.  Both $N_C$ and the number of Lagrange multipliers $h$ are linear in the volume of the unit cell while the other two loops count represent intensive quantities (e.g., energy per atom) and are expected to be (relatively) constant with unit-cell size.  Hence, the overall slave part of the calculation will scale quadratically with unit-cell volume for the current approach.

\bibliography{main}
\end{document}